\documentclass[final,onefignum,onetabnum]{siamonline190516}

\usepackage{braket,amsfonts}

\usepackage{array}

\usepackage[caption=false]{subfig}
\usepackage{pgfplots}

\newsiamthm{claim}{Claim}
\newsiamremark{remark}{Remark}
\newsiamremark{hypothesis}{Hypothesis}
\crefname{hypothesis}{Hypothesis}{Hypotheses}

\usepackage{algorithmic}

\usepackage{graphicx,epstopdf}

\Crefname{ALC@unique}{Line}{Lines}

\usepackage{amsopn}

\usepackage{xspace}
\usepackage{bold-extra}
\usepackage[most]{tcolorbox}

\colorlet{texcscolor}{blue!50!black}
\colorlet{texemcolor}{red!70!black}
\colorlet{texpreamble}{red!70!black}
\colorlet{codebackground}{black!25!white!25}

\lstdefinestyle{siamlatex}{%
	style=tcblatex,
	texcsstyle=*\color{texcscolor},
	texcsstyle=[2]\color{texemcolor},
	keywordstyle=[2]\color{texemcolor},
	moretexcs={cref,Cref,maketitle,mathcal,text,headers,email,url},
}

\tcbset{%
	colframe=black!75!white!75,
	coltitle=white,
	colback=codebackground, %
	colbacklower=white, %
	fonttitle=\bfseries,
	arc=0pt,outer arc=0pt,
	top=1pt,bottom=1pt,left=1mm,right=1mm,middle=1mm,boxsep=1mm,
	leftrule=0.3mm,rightrule=0.3mm,toprule=0.3mm,bottomrule=0.3mm,
	listing options={style=siamlatex}
}

\newtcblisting[use counter=example]{example}[2][]{%
	title={Example~\thetcbcounter: #2},#1}

\newtcbinputlisting[use counter=example]{\examplefile}[3][]{%
	title={Example~\thetcbcounter: #2},listing file={#3},#1}

\DeclareTotalTCBox{\code}{ v O{} }
{ %
	fontupper=\ttfamily\color{black},
	nobeforeafter,
	tcbox raise base,
	colback=codebackground,colframe=white,
	top=0pt,bottom=0pt,left=0mm,right=0mm,
	leftrule=0pt,rightrule=0pt,toprule=0mm,bottomrule=0mm,
	boxsep=0.5mm,
	#2}{#1}

\patchcmd\newpage{\vfil}{}{}{}
\usepackage[latin1]{inputenc}
\usepackage{amssymb}
\usepackage{graphicx}
\usepackage{cite, color}

\usepackage{cleveref}
\usepackage{esint}
\usepackage{tikz,pgfplots}
\usetikzlibrary{spy}
\usepackage{makecell}
\usetikzlibrary{pgfplots.groupplots}
\usepackage{bbm}
\usepackage{mathrsfs}
\usepackage{bm}
\usepackage{mathtools}

\def\PP{\mathbb{P}}
\def\EE{\mathbb{E}}
\def\RR{\mathbb{R}}
\DeclareRobustCommand{\argmin}{\operatorname*{argmin}}
\DeclareRobustCommand{\argmax}{\operatorname*{argmax}}

\newcommand{\RMIS}{LAIS\xspace}

\newcommand{\LSIS}{LSIS\xspace}
\newcommand{\iCEred}{iCEred\xspace}

\newcommand{\piopt}{\pi^\star_{\text{bias}}}
\newcommand{\FF}{\mathcal{F}}
\newcommand{\NN}{\mathcal{N}}
\newcommand{\VV}{\mathbb{V}}
\newcommand{\indF}{\mathbbm{1}_{\FF}}
\newcommand{\prior}{\pi_{\text{pr}}}

\newcommand{\isp}{\hat{p}_\FF^{\textsc{is}}}
\newcommand{\pibias}{\pi_{\text{bias}}}
\newcommand{\pF}{p_\FF}
\newcommand{\hpF}{\hat{p}_\FF}
\newcommand{\pMC}{\hat{p}_\FF^{\textsc{mc}}}
\newcommand{\piF}{\pi_{\FF}}
\newcommand{\piFr}{{\pi_{\FF}^{r}}}
\newcommand{\pishift}{\pi_{\textsc{lsis}}}
\newcommand{\piMIS}{\pi_{\textsc{mis}}}
\newcommand{\wMISs}{w_{\text{s-}\textsc{mis}}}
\newcommand{\wMISDM}{w_{\textsc{dm-mis}}}
\newcommand{\pMIS}{\hat{p}_\FF^{\textsc{mis}}}
\newcommand{\pSO}{\hat{p}_\FF^{\textsc{so}}}
\newcommand{\wMIS}{w_{\textsc{mis}}}

\newcommand{\pRed}{\hat{p}_\FF^{\textsc{lais}}}

\newcommand{\wshift}{w_{\textsc{lsis}}}
\newcommand{\wGauss}{w_{\textsc{gis}}}
\newcommand{\pshift}{\hat{p}_\FF^{\textsc{lsis}}}

\newcommand{\KL}{D_{\textsc{KL}}}

\newcommand{\bmu}{\bm{\mu}}
\newcommand{\bSigma}{\bm{\Sigma}}
\newcommand{\bmv}{\bm{v}}
\newcommand{\Phir}{\bm{\Phi}_{r}}
\newcommand{\Phio}{\bm{\Phi}_{\!\perp}}
\newcommand{\btheta}{\bm{\theta}}

\newcommand{\ttheta}{\tilde{\btheta}}
\newcommand{\tthetar}{\ttheta_{r}}
\newcommand{\tthetao}{\ttheta_{\!\perp}}

\newcommand{\bTheta}{\bm{\Theta}}

\newcommand{\prr}{\pi_{\text{pr}}^{r}}
\newcommand{\pro}{\pi_{\text{pr}}^{\perp}}
\newcommand{\pibiasr}{\pi_{\text{bias}}^{r}}
\newcommand{\bthetastar}{\btheta^{\star}}
\newcommand{\hatn}{\hat{\bm n}}
\newcommand{\HFIS}{{\bm{H}_{\textsc{fis}}}}
\newcommand{\HLDT}{{\bm{H}_{\textsc{ldt}}}}
\newcommand{\nCE}{N_{\!\!\:\textsc{ce}}}

\newcommand{\mur}{\bm{\mu}_r}
\newcommand{\Sigmar}{\bm{\Sigma}_r}
\newcommand{\tthetari}{\ttheta_{r,i}}
\newcommand{\tthetaoi}{\ttheta_{\!\perp,i}}

\newcommand{\cv}{\widehat{\textsc{cv}}}

\newcommand{\cvMC}{\cv^{\textsc{mc}}}

\newcommand{\cvRed}{\cv^{\textsc{lais-dm}}}
\newcommand{\cvReds}{\cv^{\textsc{lais-}\text{s}}}
\newcommand{\cvshift}{\cv^{\textsc{lsis}}}

\newcommand{\bzero}{\bm{0}}
\newcommand{\bI}{\bm{I}}
\newcommand{\simiid}{\overset{\text{i.i.d.}}{\sim}}
\newcommand{\mustar}{\bm{\mu}^\star}
\newcommand{\Sigmastar}{\bm{\Sigma}^\star}
\newcommand{\NF}{N_{F}}
\newcommand{\NdF}{N_{\nabla\! F}}

\newcommand{\rrmse}{\widehat{\textsc{rrmse}}}

\definecolor{myblue}{RGB}{0, 114, 189}
\definecolor{myorange}{RGB}{216, 83, 25}
\definecolor{myyellow}{RGB}{237, 177,  32}
\definecolor{mypink}{RGB}{239, 71,  111}
\definecolor{mypurple}{RGB}{126,   47,  142}
\definecolor{mygreen}{RGB}{119,  172,  48}
\definecolor{mycharcoal}{RGB}{38, 70, 83}
\definecolor{mycyan}{RGB}{77,  190,  238}
\definecolor{mylightred}{RGB}{255,  204,  203}

\newcommand{\bluetwo}{mycyan!80!blue}

\newcommand{\bluefour}{mycyan!40!blue}

\newcommand{\bluesix}{blue}

\newcommand{\redtwo}{mylightred!80!red}

\newcommand{\redfour}{mylightred!40!red}

\newcommand{\redsix}{red}

\pgfplotscreateplotcyclelist{my blue}{
	blue!20!black,every mark/.append
        style={solid,fill=blue!40!black},mark=*, thick\\
	blue!80!green!50!black,every mark/.append style={solid,fill=blue!80!green},mark=square*,thick\\ 
	blue!20!green!20!mycyan!50!black,every mark/.append style={solid,fill=cyan!50!blue},mark=triangle*,thick\\
	mycyan!50!black, every mark/.append style={solid,fill=mycyan},mark=pentagon*,thick\\
	mygreen!70!black,mark=otimes*,every mark/.append style={solid,
          fill=mygreen!80}, thick\\
	red!70!black,every mark/.append
        style={solid,fill=red},mark=diamond*, thick\\
}

\pgfplotscreateplotcyclelist{tsunami}{
	blue!80!green!50!black,every mark/.append style={solid,fill=blue!80!green},mark=square*,thick\\ 
	blue!80!green,,dashed, every mark/.append style={solid,fill=blue!80},mark=square*,thick\\ 
	blue!20!green!20!mycyan!50!black,every mark/.append style={solid,fill=cyan!50!blue},mark=triangle*,thick\\
	mycyan!70!black,dashed, every mark/.append style={solid,fill=cyan},mark=triangle*,thick\\
	mygreen!80!black,mark=*,every mark/.append style={solid, fill=mygreen!80},thick\\
	red!50!black,every mark/.append
	style={solid,fill=red},mark=diamond*,thick\\
	orange!70!black, dashed,every mark/.append style={solid,fill=orange!70},mark=diamond*,thick\\
}

\pgfplotscreateplotcyclelist{quadratic}{
	blue!50!black,densely dotted,every mark/.append style={solid,fill=blue!70!black},mark=square*,thick\\ 
	blue!80!green!50!black,every mark/.append style={solid,fill=blue!80!green},mark=square*,thick\\ 
	blue!80!green,,dashed, every mark/.append style={solid,fill=blue!80},mark=square*,thick\\ 
	blue!20!green!20!mycyan!30!black,densely dotted, every mark/.append style={solid,fill=cyan!50!blue!70!black},mark=triangle*,thick\\
	blue!20!green!20!mycyan!50!black,every mark/.append style={solid,fill=cyan!50!blue},mark=triangle*,thick\\
	mycyan!70!black,dashed, every mark/.append style={solid,fill=cyan},mark=triangle*,thick\\
	mygreen!80!black,mark=*,every mark/.append style={solid, fill=mygreen!80},thick\\
	red!50!black,every mark/.append
	style={solid,fill=red},mark=diamond*,thick\\
	red!30!black,densely dotted,every mark/.append style={solid,fill=red!70!black},mark=diamond*,thick\\
	orange!70!black, dashed,every mark/.append style={solid,fill=orange!70},mark=diamond*,thick\\
}

\pgfplotscreateplotcyclelist{quadratic2}{
	blue!80!green!50!black,every mark/.append style={solid,fill=blue!80!green},mark=square*,thick\\ 
	blue!20!green!20!mycyan!50!black,every mark/.append style={solid,fill=cyan!50!blue},mark=triangle*,thick\\
	mygreen!80!black,mark=*,every mark/.append style={solid, fill=mygreen!80},thick\\
	red!50!black,every mark/.append
	style={solid,fill=red},mark=diamond*,thick\\
	red!30!black,densely dotted,every mark/.append style={solid,fill=red!70!black},mark=diamond*,thick\\
	orange!70!black, dashed,every mark/.append style={solid,fill=orange!70},mark=diamond*,thick\\
}

\pgfplotscreateplotcyclelist{diffusion}{
	blue!50!black,densely dotted,every mark/.append
        style={solid,fill=blue!70!black},mark=square*,thick\\ 
	blue!80!green!50!black,every mark/.append style={solid,fill=blue!80!green},mark=square*,thick\\ 
	blue!80!green,,dashed, every mark/.append style={solid,fill=blue!80},mark=square*,thick\\ 
	blue!20!green!20!mycyan!30!black,densely dotted, every mark/.append style={solid,fill=cyan!50!blue!70!black},mark=triangle*,thick\\
	blue!20!green!20!mycyan!50!black,every mark/.append style={solid,fill=cyan!50!blue},mark=triangle*,thick\\
	mycyan!70!black,dashed, every mark/.append style={solid,fill=cyan},mark=triangle*,thick\\
	mygreen!80!black,mark=*,every mark/.append style={solid, fill=mygreen!80},thick\\
	red!50!black,every mark/.append
	style={solid,fill=red},mark=diamond*,thick\\
	orange!70!black, dashed,every mark/.append style={solid,fill=orange!70},mark=diamond*,thick\\
}

\newcommand{\data}{dataMSE/}

\title{Large deviation theory-based adaptive importance
  sampling for\\ rare events in high dimensions\thanks{Revised
    March 2023.
    \funding{Partially supported by the US National Science Foundation through
grant DMS \#1723211 and the Multidisciplinary University
Research Initiatives (MURI) Program, Office of Naval Research (ONR)
grant number N00014-19-1-2421.}}}
  \author{Shanyin Tong\thanks{Department of Applied Physics and
      Applied Mathematics, Columbia University, NY
      (\email{st3503@columbia.edu}).}
    \and Georg Stadler\thanks{Courant Institute of Mathematical Sciences, New York University, NY (\email{stadler@cims.nyu.edu}).}
  }
  \headers{LDT-based adaptive importance
  	sampling for rare events}{S.\ Tong and G.\ Stadler}

\allowdisplaybreaks

\begin{document}
	
\maketitle

\begin{abstract}
  We propose a method for the accurate estimation of rare event or
  failure probabilities for expensive-to-evaluate numerical models in
  high dimensions. The proposed approach combines ideas from large
  deviation theory and adaptive importance sampling.  The importance
  sampler uses a cross-entropy method to find an optimal Gaussian
  biasing distribution, and reuses all samples made throughout the
  process for both, the target probability estimation and for updating
  the biasing distributions. Large deviation theory is used to find a
  good initial biasing distribution through the solution of an
  optimization problem. Additionally, it is used to identify a
  low-dimensional subspace that is most informative of the rare event
  probability. This subspace is used for the cross-entropy method,
  which is known to lose efficiency in higher dimensions. The proposed
  method does not require smoothing of indicator functions nor does it
  involve numerical tuning parameters.
  We compare the method with a state-of-the-art cross-entropy-based
  importance sampling scheme using three examples:
  a high-dimensional failure probability estimation benchmark, a
  problem governed by a diffusion equation, and a
  tsunami problem governed by the time-dependent shallow
  water system in one spatial dimension.
\end{abstract}



\section{Introduction}\label{sec:intro}

Accurate probability estimation of rare, high-impact events in complex
systems with uncertainties is important in many areas. Examples of
such events include the structural failure of engineered systems,
geophysical disasters, extreme weather patterns, global disease
outbreaks, and the collapse of markets or financial systems.
Estimating the probability that such events occur is crucial, e.g.,
for hazard preparedness, for insurances companies, and
generally to make decisions about how to mitigate the effects of such
events.

To formulate the rare event probability estimation problem
mathematically, we 
consider the probability space $(\Omega, \mathscr{B}(\Omega), \PP)$
with  a probability measure $\PP$.
The uncertain parameter $\btheta$ is modeled as a random vector on
$\Omega$ as $\btheta: \Omega\to\bTheta\subseteq \RR^n$ with $\bTheta$
is a measurable set. We assume that the distribution of $\btheta$ has a
probability density $\prior$, which we also call the \emph{prior}.
Further, we consider a (possibly complicated) \emph{parameter-to-event
map} $F: \bTheta \to \RR$, which maps the random parameter vector
$\btheta$ to the quantity of interest or event outcome $F(\btheta)$.  We are
interested in events $\FF \coloneqq\{ \btheta : F(\btheta) \geq z\}$,
where $z\in \RR$ is a given threshold that captures the extremeness or
maximal allowance of the event. 
Our aim is to estimate the
rare event probability, also called probability of failure
\begin{equation}
\label{eq:prob-fail}
\pF \coloneqq \PP[\FF] = \int_{\bTheta} \indF(\btheta) \prior(\btheta) d\btheta = \EE_{\prior} [\indF(\btheta)],
\end{equation}
where $\indF: \bTheta\to \{0, 1\}$ is the indicator function, i.e.,
$\indF(\btheta)=1$ if $\btheta \in \FF$, and $\indF(\btheta)=0$ else.

\subsection{Challenges and related literature}
Since the random parameter $\btheta$ may be defined over a high-dimensional
space and the map $F$ may be complicated, e.g., involve the solution
of a differential equation, analytic calculation of
\cref{eq:prob-fail} is generally intractable.
Monte Carlo methods are the standard
approach to study complex systems that include
uncertainty. 
The standard Monte Carlo estimator \cite{metropolis1949monte,liu2008monte} for $\pF$ is 
$ %
\pMC \coloneqq \frac{1}{N} \sum_{i=1}^N\indF(\btheta_i), 
$ %
where $\{\btheta_i\}_{i=1}^N$ are random draws from $\prior$. This estimator is
unbiased since $\EE_{\prior}[\pMC] = \pF$, and its variance is
\begin{equation}
  \label{eq:var-MC}
\VV_{\!\prior}[\pMC] = \frac{1}{N} \VV_{\!\prior}[\indF(\btheta)] = \frac{1}{N} (\pF -\pF^2).
\end{equation}
The coefficient of variation, a measure for the accuracy of the
estimator, is defined as the ratio between the
standard deviation and the expectation, i.e.,
\begin{equation}
\label{eq:cv-MC}
\cvMC = \frac{\sqrt{\VV_{\!\prior}[\pMC]}}{\EE_{\prior}[\pMC]} = \frac{1}{\sqrt{N}} \frac{\sqrt{\pF-\pF^2}}{\pF} \approx \frac{1}{\sqrt{N}} \frac{1}{\sqrt{\pF}}.
\end{equation}
For linear or moderately nonlinear maps $F$, the target probability $\pF$ decays exponentially as $z$ is increased \cite{tong2021extreme}. Thus,
to maintain the same error level, the sample size $N$ would have to increase
exponentially, making the Monte Carlo estimation \cref{eq:cv-MC}
very costly. Thus, these methods are insufficient to explore
probabilities of rare events, which are typically
in the range of $10^{-3}$--$10^{-10}$.

Various approaches to overcome the challenges arising in rare event
probability estimation have been proposed, including 
parameter space decomposition, sequential sampling and multilevel
splitting
\cite{ditlevsen1990general,ullmann2015multilevel,
  wagner2020multilevel, au2001estimation, cerou2012sequential,sapsis2018new}. In this work, we
focus on importance sampling and thus restrict our discussion to the
relevant literature.
Importance sampling (IS) \cite{kahn1953methods,
  bucklew2007introduction,liu2008monte} decreases the required number
of samples by choosing a biasing proposal distributions that reduces the
variance of the estimator. The key to applying IS for rare
event sampling is to find a good biasing density. While the form of
the optimal biasing density is known, it cannot be used directly
because of the unknown normalizing constant.
One approach to find a good biasing density are
cross-entropy (CE) methods, which find proposals by minimizing the Kullback-Leibler (KL)
divergence between a parametrized density (usually from an exponential
family) and the optimal biasing density \cite{de2005tutorial,
  geyer2019cross}.  However, CE methods are typically
limited to low-dimensional problems. This is because finding
an optimal density is itself based on sample
evaluations, and the number of required samples increases
with the dimension. Recent work combines dimension reduction ideas
from Bayesian inverse problems \cite{zahm2018certified} with CE to
overcome this challenge \cite{uribe2021cross}. A related approach is
followed in  \cite{wahalbimc}, where the authors also borrow ideas from
Bayesian inference to define a maximum a posteriori (MAP)
point in parameter space and to construct Gaussian distributions to be
used as proposal densities for IS sampling.

The samples used to find parameterized CE proposals should be
drawn from the parameter region around the rare event
set.
Finding these regions is difficult since typically one does not have
much prior knowledge about which parameters lead to rare events.
Thus, adaptive procedures are used, in which one
starts from an initial proposal, and iteratively updates the
parametric proposal using samples and evaluations from the previous
step. These procedures are called adaptive importance sampling (AIS)
methods \cite{bugallo2017adaptive,paananen2021implicitly}, and CE is
one method used in AIS for updating biasing densities.

Taking the perspective of large deviation theory (LDT)
has also proven useful to estimate rare event probabilities
in system with random components
\cite{dematteis2019extreme, tong2021extreme} and to sample rare and
large fluctuations in non-equilibrium stochastic PDE systems \cite{ebener2019instanton}.
In this approach, one solves an optimization problem to find the most
important point (also called instanton) in the rare event
set. Knowledge of this point provides asymptotic information about small
probabilities and may help to guide IS \cite{tong2021extreme,ebener2019instanton}. 
Probability estimation of rare events is also related to
reliability analysis in engineering \cite{ditlevsen1996structural,
  lemaire2013structural}. Methods used in this context are based on
the point with largest probability density (typically of a Gaussian
distribution), combined with set approximations based on Taylor
expansions called first and second order reliability methods (FORM and
SORM) \cite{du2001most,rackwitz2001reliability,schueller1987critical}.
IS methods based on the most probable point have also been proposed
\cite{kahn1953methods, schueller1987critical}.

\subsection{Approach}
The approach we propose in this paper combines several of the ideas
reviewed above. It uses large deviation theory and a cross-entropy
method to identify an optimal parametric biasing density for importance sampling.
Since we target problems with high-dimensional parameter space,
cross-entropy approaches can become inefficient. Thus, we use
cross-entropy only in the subspace that is most important for
estimating the rare event probability. To identify this
subspace, we use derivative information of the parameter-to-event map
at the large deviation optimizer. This step borrows geometric
approximation ideas for rare
event set from reliability analysis.

Overall, this results in an adaptive importance sampling scheme for
rare event probability estimation in high dimensions. The method first
solves a constrained optimization problem to find
the LDT optimizer in random space. Using first- and second-order derivative
information of the parameter-to-event map at the LDT optimizer, we
identify a low-dimensional subspace in which we perform a cross-entropy method
to identify an optimal Gaussian biasing density for importance
sampling. Repeating this procedure, we obtain an adaptive multiple
importance sampling method that reuses all samples taken in the process.

\subsection{Contributions and limitations}
The main contributions made in this paper are as follows:
(1) We propose a method for rare event probability estimation that
does not require smoothing of the indicator function in
\cref{eq:prob-fail}, and has no critical numerical parameters that
must be tuned.
(2) Our method requires derivatives of the parameter-to-event map
only in the initial phase consisting of LDT optimization and
identification of the low-dimensional subspace informing the failure
probability. The cross-entropy adaptive sampling procedure is
derivative-free.
(3) After the initial LDT optimization, all evaluations of the
parameter-to-event map in the adaptive importance sampling are used in the final probability estimation.

Our approach also has some limitations.
(1) We assume that the prior is a multivariate Gaussian or the
push-forward of a multivariate Gaussian. Large deviation theory allows
for more general priors, but for non-Gaussian priors it may not be straightforward to identify a
low-dimensional failure-informing subspace.
(2) The proposed method requires solution of an optimization problem
which is assumed to have a unique minimizer.
(3) The LDT-informed subspace we use does not have guaranteed
certification properties. However, the choice of subspace does not
affect the correctness of the solution as it is only used in the construction of the
biasing density for importance sampling.
(4) For one variant of the proposed adaptive importance
sampler, one cannot prove that it has zero bias. However, in our
experiments (and in related experiments in the literature), one does
not observe significant bias.

\section{Preliminaries}
In this section, we summarize basic properties of importance sampling
(IS) and cross-entropy (CE) methods, which provide a parametric approximation of 
the
optimal biasing density for IS. We also highlight the
limitations these methods have for rare events and high-dimensional
parameter spaces, which will then be addressed by the
methods presented in \cref{sec:LDT}.

In this work, we
assume that the random vectors $\btheta$ are distributed according to
a multivariate Gaussian distribution, i.e., $\prior$ is a Gaussian
density. This is a reasonable assumption in many applications. The
methods we propose can easily be generalized to distributions that are
push-forwards of multivariate Gaussians under a nonlinear map such as
a Rosenblatt or Nataf transformation
\cite{lemaire2013structural}, or to Gaussian
mixtures
\cite{bilmes1998gentle}. Since any finite-dimensional multivariate Gaussian can
 be linearly mapped to a standard multivariate normal $\NN(\bzero,
\bI_n)$, we assume, without loss of generality, that
$\prior\sim\NN(\bzero, \bI_n)$ for the remainder of this paper.
We remark that in the context of failure probability estimation in
engineering, it is common to define the \emph{limit-state function
(LSF)} as $g(\btheta) \coloneqq z-F(\btheta)$. Then, the
failure set can be equivalently be written as $\FF =\{ \btheta :
g(\btheta) \leq 0 \}$.

\subsection{Importance sampling}\label{sec:IS}
The ineffectiveness of standard MC for rare event probability
estimation is due to the large number of samples $\btheta_i$ that fall
outside of $\FF$, which implies that $\indF(\btheta_i) = 0$.  The idea
of importance sampling (IS) is to sample from a \emph{proposal/biasing
density} $\pibias$ that has high density in the failure set $\FF$
rather than from the prior distribution. This can substantially
increase the information provided by each sample evaluation
\cite{bucklew2007introduction,kahn1953methods,liu2008monte}.
The IS estimator is defined as 
\begin{equation*}
\isp \coloneqq \frac{1}{N}\sum_{i=1}^N \indF(\btheta_i) w(\btheta_i), \; \text{with } w(\btheta_i) = \frac{\prior(\btheta_i)}{\pibias(\btheta_i)},
\end{equation*}
where the samples $\{\btheta_i\}_{i=1}^N\overset{\text{i.i.d.}}{\sim}
\pibias$ are independent and identically distributed samples from a
distribution with density $\pibias$. 
Specially, when the bias density is a Gaussian with mean $\bmu$ and covariance $\bSigma$, the IS weight has the explicit formula
\begin{equation}
\label{eq:w-GIS}
\wGauss(\btheta; \bmu,  \bSigma) \coloneqq (\det
\bSigma)^{\frac{1}{2}}\exp\left(-\frac{1}{2}
\|\btheta\|^2+\frac{1}{2}\|\btheta -
\bmu\|^2_{\bSigma^{-1} }\right).
\end{equation}
The IS estimator $\isp$  is unbiased, i.e., $\EE_{\pibias}[\isp]=\pF$, provided that the support of the biasing density $\pibias$ contains $\FF$, which is true for Gaussian proposals.
Its variance is
\begin{equation}
\label{eq:IS-var}
\VV_{\!\pibias}[\isp] = \frac{1}{N} \VV_{\!\pibias} [\indF(\btheta) w(\btheta)] =\frac{1}{N} (\EE_{\pibias} [\indF(\btheta) w^2(\btheta)] -\pF^2).
\end{equation}
The minimizer of $\VV_{\!\pibias}[\isp]$ over all possible biasing densities
$\pibias$ is the optimal biasing density %
\begin{equation}
\label{eq:opt-bias}
\piF(\btheta)\coloneqq\frac{\indF(\btheta)\prior(\btheta)}{\int \indF(\btheta)\prior(\btheta) d\btheta}
= \frac{1}{\pF} \indF(\btheta)\prior(\btheta) ,
\end{equation}
which provides a zero-variance IS estimator, i.e.,
$\VV_{\!\pibias}[\isp] = 0$ when $\pibias=\piF$. However, it is
impractical to use \cref{eq:opt-bias} for IS, because the
normalizing constant $\pF$ is unknown; indeed, it is the target of the
estimation problem.  Although it is impossible to evaluate and sample directly from
$\piF$, this still provides a guide on how to choose an IS biasing
distribution. Next, we summarize cross-entropy (CE) methods
for obtaining approximations of $\piF$, from which we can draw samples
more easily.

\subsection{Cross-entropy method (CE)}\label{sec:CE}
As discussed above, the optimal biasing density $\piF$ in
\cref{eq:opt-bias} is not available. The goal of cross-entropy methods
\cite{de2005tutorial, rubinstein1997optimization} is to find approximations for
optimal biasing densities. CE methods approximate a target
distribution $\piopt$ (such as $\piF$ in our case) with a parametric
biasing density $\pibias(\btheta; \bmv)$ with reference parameter
$\bmv$ by choosing from a family of densities $\{ \pibias(\cdot; \bmv)
\}$ the one that minimizes the Kullback-Leibler (KL) divergence
$\KL(\piopt||\pibias(\btheta; \bmv))$. This approximated distribution
is then used to sample and evaluate the IS estimator. For Gaussian
parametric biasing densities, the parameters $\bmv$ are the mean
$\bmu$ and covariance matrix $\bSigma$.  From the definition of the KL
divergence,
\begin{equation}
\KL(\piopt||\pibias(\btheta; \bmv)) = \EE_{\piopt} \ln \left( \frac{\piopt(\btheta)}{\pibias(\btheta;\bmv)}\right) = \EE_{\piopt} \ln \piopt(\btheta) - \EE_{\piopt}\ln \pibias(\btheta;\bmv).
\end{equation}
Since the first term on the right hand side above does not depend on $\bmv$,
minimizing the KL divergence is equivalent to the following problem:
\begin{equation}\label{eq:CE}
\begin{aligned}
\bmv^\star = \argmax_{\bmv} \EE_{\piopt} [\ln \pibias(\btheta; \bmv)]
= \argmax_{\bmv} \EE_{\prior} [\ln \pibias(\btheta; \bmv)\indF(\btheta)].
\end{aligned} 
\end{equation}
The expectation can be rewritten 
under the measure of
another biasing density $\pibias'(\cdot)$ and the corresponding IS weight $w'(\cdot)$, thus the maximization
problem is equivalent to 
\begin{equation}\label{eq:CE-IS}
\bmv^\star = \argmax_{\bmv} \EE_{\pibias'}[\ln \pibias(\btheta; \bmv)\indF(\btheta) w'(\btheta)], \; \text{with } w'(\btheta) = \frac{\prior(\btheta)}{\pibias'(\btheta)}.
\end{equation}
CE methods choose $\pibias'(\cdot) = \pibias(\cdot; \bmv')$ from the same density family, start with an initial choice of $\bmv'$ and iteratively update the reference parameter $\bmv^\star$ using \cref{eq:CE-IS}.

For a Gaussian density $\pibias(\btheta; [\bmu, \bSigma]) =
(2\pi)^{-\frac{n}{2}}(\det
\bSigma)^{-\frac{1}{2}}\exp(-\frac{1}{2}\|\btheta
-\bmu\|^2_{{\bSigma}^{-1} })$, the optimal mean and covariance can be
computed analytically from \cref{eq:CE-IS}:
\begin{equation}
\label{eq:mu-cov-IS}
\mustar = \frac{\EE_{\pibias'}  [\indF(\btheta) w'(\btheta) \btheta] }{\EE_{\pibias'} [ \indF(\btheta) w'(\btheta)] }, \; \Sigmastar = \frac{\EE_{\pibias'}  [\indF(\btheta) w'(\btheta) (\btheta- \mustar)(\btheta- \mustar)^\top] }{\EE_{\pibias'}  [\indF(\btheta) w'(\btheta)] }.
\end{equation}
Using samples  $\{\btheta_i\}_{i=1}^N\simiid \pibias'$,
the optimal mean and covariance
$\bmv^\star$ can be calculated as \cite{geyer2019cross}:
\begin{equation}\label{eq:mu-cov}
\mustar = \frac{\sum_{i=1}^N \indF(\btheta_i) w'(\btheta_i) \btheta_i }{\sum_{i=1}^N \indF(\btheta_i) w'(\btheta_i) }, \;\Sigmastar = \frac{\sum_{i=1}^N \indF(\btheta_i) w'(\btheta_i) (\btheta_i-\mustar)(\btheta_i-\mustar)^\top }{\sum_{i=1}^N \indF(\btheta_i) w'(\btheta_i) }.
\end{equation}
From the above derivation, we identify three main challenges for the
CE method when applied to high-dimensional problems with
expensive-to-evaluate parameter-to-event maps:
\begin{enumerate}
\item[(1)]\label{item:1} To obtain useful estimates \cref{eq:mu-cov} of $\mustar$ and
  $\Sigmastar$ with a moderate number of samples $N$, a good biasing
  density $\pibias'$ is needed, for which a large fraction of
  evaluations $\indF(\btheta_i)$ are nonzero and the weights
  $w'(\btheta_i)$ are large.
\item[(2)] When $\btheta$ is high-dimensional, a large number of samples are
  required to estimate $\mustar$ and $\Sigmastar$ in
  \cref{eq:mu-cov}, even if a good biasing density $\pibias'$ is
  available.
\item[(3)] Samples $\{\btheta_i\}_{i=1}^N\simiid \pibias'$ used to
  (iteratively) find optimal parameters $\bmv^\star$ are typically not
  used in the final estimator, which wastes expensive evaluations of
  $\FF$. This also separates the method into a step to find
  the biasing density, and into the IS step to estimate the rare event
  probability. Thus, it requires to choose when to switch from one
  step to the other.
\end{enumerate}

In \cite{uribe2021cross}, the authors tackle some of the above
challenges. They address (1) using a sequence of smoothed
approximations of $\indF$ with a smoothing parameter $s$ to increase
the ratio of nonzero evaluations, and construct a failure-informed
subspace to reduce the dimension of the problem, addressing (2). In
particular, they start with the prior distribution as $\pibias'$ and
large $s$ (i.e., a rather smooth approximation), obtain the
low-dimensional subspace using gradient information of a smoothed
version of $\indF$ (details in \cref{sec:FIS}) and apply the CE method
in this subspace. Then, they reduce $s$ and use this newly-obtained
proposal as $\pibias'$ to update the subspace and the CE mean and
covariance. This step is repeated until a good IS biasing density is
found. This requires a substantial number of evaluations of
$F$ and $\nabla F$ which are not used for the final probability
computation. Additionally, the
cost of this procedure increases with the rareness of the event.
In the next section, we use ideas from large deviation theory to find
an initial biasing density using optimization, addressing (1). We
then use local derivative information at the optimizing point to
identify a low-dimensional subspace, addressing (2). Since we do not
require smoothing, our method involves few parameters and uses
all evaluations of $F$ in the sampling step for the probability estimation by 
employing a
multiple IS method, addressing (3).

\section{LDT-based adaptive importance sampling}
In this section, we show how large deviation theory can provide
an initial biasing density for adaptive importance sampling
(\cref{sec:LDT}). Moreover, LDT also
provides dimension reduction information required to apply CE
methods in a subspace only (\cref{sec:dim-red}).
Combining these ideas, we propose a multiple IS algorithm for the
target probability that reuses all evaluations of $F$ in the sampling step and
thus seamlessly merges the parametric biasing density optimization and the probability
estimation phases.

\subsection{Large deviation theory (LDT)}\label{sec:LDT}
Large deviation theory is a classical theory on the
asymptotic behavior of sequences of probability
distributions \cite{varadhan1984large,dembo1998large}.
Recent studies adapt LDT to systems with random parameters
and build connections between extreme event probability estimation and
constrained optimization \cite{tong2021extreme,
  dematteis2019extreme, tong2022optimization}.
LDT states that the least-unlikely point $\bthetastar$ in the event set
$\FF$ carries important information about $\pF$. We refer to
$\bthetastar$ as the LDT optimizer, and it is the solution
of the constrained optimization problem
\begin{equation}\label{eq:LDT-opt}
\bthetastar \coloneqq \argmin_{\btheta\in\FF} I(\btheta),
\end{equation}
where the rate function $I$ depends only on $\prior$.
For the standard normal case $\NN(\bzero, \bI_n)$, the rate function
is simply
$
I(\btheta) = \frac{1}{2} \|\btheta\|^2$. Then, LDT shows that under
reasonable regularity conditions on $F$, the probability $\pF$ is
log-asymptotic to $\exp(-I(\bthetastar))$, i.e., $\ln\pF\rightarrow
{-I(\bthetastar)}$ as $z\to\infty$. Using this directly is
limited due to the occurrence of the logarithm and due to its
asymptotic nature, i.e., it only holds in the limit as the probability
tends to zero. Significant research has been performed to improve this
log-asymptotic estimate through estimation of prefactors either in
the context of stochastic differential equations
\cite{ebener2019instanton,grafke2019numerical,schorlepp2022spontaneous}  
or systems
with uncertain parameters \cite{dematteis2019extreme, tong2021extreme}.

The LDT optimizer $\bthetastar$ induces a proposal for an IS method that is centered at
$\bthetastar$. That proposal is $\NN(\bthetastar, \bI_n)$, and the
corresponding 
LDT-based shifted importance sampling (\LSIS) estimator is
\begin{equation}
\label{eq:est-shift}
\pshift := \frac{1}{N}\sum_{i=1}^N \indF(\btheta_i)\wshift(\btheta_i), \; \text{with } \wshift(\btheta_i) \coloneqq \wGauss(\btheta_i;\bthetastar, \bI_n),
\end{equation}
where $\{\btheta_i\}_{i=1}^N\overset{\text{i.i.d.}}{\sim} \NN(\bthetastar, \bI_n)$
are samples from the shifted prior, and $\wGauss$ is defined in \cref{eq:w-GIS}.
For linear or moderately nonlinear $F$ and for extreme events (large $z$ and small $\pF$), the error of the \LSIS estimator
is reduced by an exponential term compared with standard Monte Carlo methods
\cite{tong2021extreme}. For severely nonlinear problems in high
dimensions, %
this method might underestimate the probability due to a large number of zero evaluations of the indicator function at the generated samples \cite{katafygiotis2008geometric}.
In the context of SDEs, the
optimizer \cref{eq:LDT-opt} is also called \emph{instanton}, and can be
used to form an alternative SDE which
allows to sample, e.g., 
large fluctuations in non-equilibrium systems
\cite{ebener2019instanton}. The resulting instanton-based importance
sampling method is analogous to \LSIS.
\LSIS
serves as a good initial choice of the biasing density $\pibias' =\pishift$,
acting as alternative to using smooth approximations of the indicator function as in
\cite{uribe2021cross} to avoid a large number of zero evaluations of $\indF$.

\subsection{Dimension reduction}\label{sec:dim-red}
As discussed in \cref{sec:CE}, the effectiveness of cross-entropy
methods depends crucially on the choice of the biasing distribution
used to compute the reference parameters. As can be seen in
\cref{eq:mu-cov}, each sample updates the
covariance matrix approximation by at most a rank-1
matrix. Since the covariance must be positive and the approximations of
the mean and covariance sufficiently accurate, one typically needs
$\nCE\gg n$ samples, where $n$ is the dimension of the parameter space
and $\nCE$ is the sample size used in the CE method for finding optimal reference parameters.
When $n$ is large and evaluation of $F$ is
costly, using CE might thus not be practical.
However, rare event probability estimation problems often have
intrinsic low-dimensional structure, resulting from
properties of the parameter-to-event map and correlation of the random
parameters.  We exploit such structure by tailoring the
biasing distribution $\pibias$ in the low-dimensional subspace
using the CE method, thus drastically reducing the number of required
samples. In the next section, we use the relationship between rare
event probability estimation and Bayesian inference, allowing the
transfer of dimension reduction ideas from the latter to the former.
In \cref{sec:LDT-low}, we use ideas from LDT to identify an appropriate
low-dimensional space that is informative for rare events, and in \cref{sec:CE-lowdim} we
show how to use CE in that subspace.

\subsubsection{Low-dimensional subspace informing rare event probability}\label{sec:FIS}
Assume there is a low-dimensional subspace of the random parameter space that
is important for the rare event probability estimation. Before
discussing how to find such a space, we show how, once available, it can
be used. We assume an
orthonormal basis of $\mathbb R^n$ such that this subspace is spanned
by the columns of $\Phir\in\RR^{n\times r}$, and the columns of $\Phio
\in \RR^{n\times (n-r)}$ are the remaining orthonormal basis vectors. This induces
the orthogonal decomposition
$\btheta = \Phir\Phir^\top \btheta + \Phio\Phio^\top \btheta =\Phir
\tthetar + \Phio \tthetao$, where
$\ttheta=[\tthetar,\tthetao]\in\RR^n$ has the components
$\tthetar\coloneqq\Phir^\top \btheta \in \RR^r$ and $\tthetao\coloneqq\Phio^\top
\btheta \in \RR^{n-r}$ are the coordinates of $\btheta$ for the basis
$\bm\Phi\coloneqq[\Phir, \Phio]$.

Partially motivated by dimension reduction ideas from Bayesian inverse problem,
i.e., considering $\piF$ as a posterior density and $\indF(\btheta)$
as the likelihood, the optimal low-dimensional density $\piFr$ for a
subspace with basis $\Phir$ can be obtained by minimizing the KL
divergence $\KL(\piF||\piFr)$, resulting in,
\cite{uribe2021cross,zahm2018certified}
\begin{equation}
\label{eq:approx-bias}
\piFr(\btheta) \propto \EE_{\prior} [\indF(\btheta)|\Phir\Phir^\top \btheta] \prior (\btheta),
\end{equation}
where the conditional expectation is defined as
\begin{equation*}
\EE_{\prior} [\indF(\btheta)|\Phir\Phir^\top \btheta] \coloneqq \int_{\RR^{n-r}} \indF(\Phir\Phir^\top \btheta + \Phio \tthetao)\frac{\prior(\Phir\Phir^\top \btheta + \Phio \tthetao)}{\int_{\RR^{n-r}} \prior(\Phir\Phir^\top \btheta + \Phio \tthetao') d\tthetao'} d\tthetao.
\end{equation*}
The remaining question is how to choose the subspace and its basis
$\Phir$. In \cite{uribe2021cross}, the authors use the
failure-informed subspace (FIS), which is computed from the
generalized eigenvalues of the matrix
\begin{equation}
\label{eq:H-FIS}
\HFIS \coloneqq \EE_{\piF} \nabla \ln f(\btheta;s) \nabla \ln f(\btheta;s)^\top,
\end{equation}
where $f(\btheta;s)$ is a smooth approximation of $\indF (\btheta)$,
e.g., $f(\btheta;s)\coloneqq
\frac{1}{2}\left[1+\text{tanh}\left({(F(\btheta)-z)}/{s}\right)\right]$. The
dominant generalized
eigenvectors of $\HFIS$ form the columns of $\Phir$.
This approach adopts dimension
reduction methods originally developed for Bayesian inverse problems in
\cite{zahm2018certified}, for which certified error bound for
$\KL(\piF||\piFr)$ are available.  The corresponding
cross-entropy-based IS with failure-informed
dimension reduction is termed \iCEred in \cite{uribe2021cross}. We will
compare the performance of this method  to the
approach we propose in this paper.

Alternative dimension reduction methods, namely local and global
likelihood-informed subspaces (LIS) are proposed in the context of
Bayesian inverse problems in \cite{cui2014likelihood}.  Motivated by the Bayesian
linear-Gaussian model, these approaches use the subspace spanned by
the dominant eigenvectors of the Gauss-Newton Hessian
of the negative log-likelihood (either at a point or integrated over
the posterior $\piF$). These LIS projectors do not come with certified
error bounds for the resulting posterior approximation. 
Both the certified dimension reduction using the dominant eigenspace
of the matrix \cref{eq:H-FIS} and the global LIS methodologies perform
comparably well in applications for nonlinear Bayesian inverse problems
\cite{zahm2018certified}.

\subsubsection{LDT-informed subspace}\label{sec:LDT-low}
Dimension reduction based on \cref{eq:H-FIS} and the global LIS method
both require sampling over the posterior $\piF$ and evaluation of
derivatives to build the low-rank projector. We would like to limit
the number of evaluations of $F$ and its derivatives as these
typically dominate the computational cost. For that purpose, we use
geometric information at the LDT optimizer (see \cref{sec:LDT}) and a
local LIS approach to identify a low-dimensional subspace,
building on the result that one can asymptotically approximate the rare event probability by using
local derivative information at the LDT optimizer.
In the engineering literature,
a similar perspective is referred to as the second-order
reliability method (SORM) \cite{du2001most}. For
reference, we summarize a result from \cite{tong2021extreme} here.

\begin{theorem}[Second-order approximation]\label{thm:SO}
	Let $\btheta\sim\NN(\bzero,\bI_n)$ and assume that $F$ is
        twice continuously differentiable. Denote by $\bthetastar$ the
        optimizer of \cref{eq:LDT-opt}, and assume that the minimizer
        $\bthetastar$ satisfies a second-order sufficient optimality
        condition, i.e., $\bI_n -\lambda \nabla^2
        F(\bthetastar)$ is positive definite, where $\lambda\coloneqq\|\nabla
        I(\bthetastar)\|/\|\nabla F(\bthetastar)\|$. Denote the
        normal direction at $\bthetastar$ as $\hatn\coloneqq\nabla
        F(\bthetastar)/\|\nabla F(\bthetastar)\|$. Then, an
        approximation $\pSO$ of $\pF$ using a second-order
        approximation of the rare event set can be computed as
	\begin{equation}
	\label{eq:pSO}
	\pSO \coloneqq \dfrac{(2\pi)^{-\frac 12}}{\sqrt{2I(\bthetastar)}} \prod_{i=1}^{n}\left[1 -\lambda
	\lambda_i\left(\HLDT \right)  \right]^{-\frac{1}{2}} \exp(-I(\bthetastar)),
	\end{equation}
	where
        \begin{equation}\label{eq:HLDT}
        \HLDT\coloneqq{(\bI_n-\hatn\hatn^\top)}\nabla^2 F(\bthetastar){
          (\bI_n-\hatn\hatn^\top)}
        \end{equation}
        is the Hessian of $F$ at $\bthetastar$ projected
        onto the hyperplane orthogonal to $\hatn$, and
        $\lambda_i(\HLDT)$ is its $i$-th
        eigenvalue.  Moreover, this approximation is asymptotic to
        $\pF$, i.e.,
        ${\pSO} \to {\pF}$ as $z\to\infty$.
\end{theorem}
For fixed $\bthetastar$, we observe that the values contributing most
to the product in \cref{eq:pSO} are $\lambda$, which is related to
the gradient norm $\|\nabla F(\bthetastar)\|$, and the dominating
eigenvalues of $\HLDT$.  This suggests that $\hatn$ and the
corresponding eigenvectors of $\HLDT$ are important directions for the rare event
probability.

As we will show next, this observation can be confirmed by using the relation between rare event probability
estimation and Bayesian inverse problems and following similar steps
as in \cite{uribe2021cross}.
In Bayesian inference, dimension reduction is based on
the goal to
identify the subspace that is most important for the update from the
prior to the posterior.
As in \cite{uribe2021cross}, to avoid the discontinuity of to the
indicator function $\indF(\btheta)$, we introduce a smooth
approximation
with smoothing parameter $s>0$,
\begin{equation}\label{eq:smooth-cdf}
	f(\btheta;s)\coloneqq\frac{1}{2}\left[1+\text{tanh}\left(\frac{F(\btheta)-z}{s}\right)\right], \text{ where } \lim_{s\to 0} f(\btheta;s )=\indF(\btheta).
\end{equation}
Thus, the negative log-likelihood is approximated as
$-\ln f(\btheta; s) =\ln2 -\ln\left[1+\text{tanh}\left({(F(\btheta)-z)}/{s}\right)\right].$
We now compute derivatives with respect to
$\btheta$. Using the notation 
$\alpha (\btheta;s)\coloneqq 1- \text{tanh}\left({(F(\btheta)-z)}/{s}\right) $,
the
gradient and Hessian are
\begin{align}
\nabla(-\ln f(\btheta; s) )&= -\frac{\alpha(\btheta;s)}{s} {\nabla
  F(\btheta)} \label{eq:dlog-smooth}\\
\nabla^2(-\ln f(\btheta; s) )
&= %
 -\frac{\alpha(\btheta;s)}{s} {\nabla^2 F(\btheta)} %
 + \frac{2\alpha(\btheta;s)-\alpha^2(\btheta;s)}{s^2} 
  {\nabla F(\btheta) \nabla
   F(\btheta)^\top}. \label{eq:ddlog-smooth}
\end{align}
In \cite{uribe2021cross}, the authors derive a low-dimensional
failure-informed subspace by sampling the gradients
\cref{eq:dlog-smooth} with fixed regularization parameter $s$ to
approximate  \cref{eq:H-FIS}.
They use the
dominating eigenvalue directions of the resulting second moment matrix
to compute the low-dimensional, failure-informed subspace and repeat
this procedure for decreasing $s>0$.  An alternative approach to
dimension reduction is using the likelihood-informed subspace (LIS),
which is spanned by the dominant eigenvectors of the Gauss-Newton Hessian  of the negative
log-likelihood
\cite{cui2014likelihood}. 
Averaging of Hessians over the posterior density leads
to the global LIS method, while using the Hessian at a point amounts
to the local LIS method. We use the latter here and evaluate the
Hessian at the LDT optimizer $\bthetastar$. This is also motivated by
large deviation theory, which shows that the probability measure
concentrates around $\bthetastar$ as $z\to\infty$.

To study the dominating eigenvalues of the Hessian
\cref{eq:ddlog-smooth} for $\btheta=\bthetastar$ and as $s\to 0$, note that
$\alpha(\bthetastar,s)=1$ 
independent of $s$. Thus, for
sufficiently small $s$, the first term in \cref{eq:ddlog-smooth} is
dominated by the second. As a consequence, $\nabla F(\bthetastar)$ is
a dominating eigenvector direction of the Hessian \cref{eq:ddlog-smooth} and
should thus be in the local LIS. Let us now consider directions that
are orthogonal to $\nabla F(\bthetastar)$. These directions are in the
nullspace of the second term in \cref{eq:ddlog-smooth}, and thus the
dominating directions are the dominating eigenvectors of
$\nabla^2 F(\bthetastar)$ in the hyperplane orthogonal to $\hatn$. These can be computed as dominant eigenvectors
of $\HLDT$ defined in \cref{eq:HLDT}, i.e., the same information
as appears in \cref{thm:SO}.

To summarize, the local LIS is spanned by $\nabla F (\bthetastar)$ (or
equivalently, its normalized version $\hatn$) and the dominating
eigenvectors of $\HLDT$, which are both independent of $s$.
Note that this definition
only requires derivative
information of $F$ at the LDT optimizer $\bthetastar$.
Using the notation from \cref{sec:FIS}, the subspace basis vectors form the columns of
$\Phir$. The first column is the normal direction $\hatn$, and the other columns are
the dominating eigenvectors of $\HLDT$.  The decay of the eigenvalues
of $\HLDT$ can be used together with the value of $\lambda$ to choose the rank $r$ of the low-dimensional
subspace. For example, we can choose $r$ such that
\begin{equation}\label{eq:epsilon-r}
  \lambda\max_{i>r}|\lambda_i(\HLDT)| \leq\epsilon,
\end{equation}
where $\epsilon$ is a given threshold.
In practice, we do not build the full matrix
$\HLDT$, but instead use its application to vectors and 
Lanczos or randomized SVD methods to find the dominating eigenvectors
\cite{CullumWilloughby85,HalkoMartinssonTropp11}. The number of
required matrix-vector applications for these methods is typically
only moderately larger than the number of dominant eigenvalues.

\subsubsection{CE in low-dimensional subspace}\label{sec:CE-lowdim}
We now show how to use CE methods for approximating the optimal
low-dimensional biasing density $\piFr$. This section largely
follows \cite{uribe2021cross} but we provide a summary for completeness.
Using 
\cref{eq:approx-bias}, we define the target density as
\begin{equation}
\label{eq:opt-bias-low}
\piopt(\ttheta) \coloneqq \piFr(\btheta)  \propto \EE_{\prior} [\indF(\btheta)|\Phir \tthetar]\prr(\tthetar) \pro(\tthetao),
\end{equation}
where $\btheta=\bm\Phi\ttheta$ and we use that the prior distribution
can be factorized as $\prior (\btheta) =\prr(\tthetar) \pro(\tthetao)$.
Similarly, the parametric biasing density is defined as
\begin{equation}
\label{eq:para-bias-low}
\pibias(\ttheta;\bmv_r) \coloneqq \pibiasr(\tthetar; \bmv_r)\pro(\tthetao),
\end{equation}
where $\bmv_r$ represents the mean and covariance in the $r$-dimensional subspace.

To apply CE to minimize $\KL(\piopt||\pibias)$ using
\cref{eq:opt-bias-low} and \cref{eq:para-bias-low}, we solve
the following problem \cite{uribe2021cross},
\begin{equation}\label{eq:CE-low}
\begin{aligned}
\bmv_r^\star =& \argmax_{v_r} \EE_{\piopt} [\ln \pibias(\ttheta; \bmv_r)] = \argmax_{v_r} \EE_{\piopt} [\ln \pibiasr(\tthetar; \bmv_r) ]\\
=& \argmax_{v_r} \int_{\RR^r} [\ln \pibiasr(\tthetar; \bmv_r) ] \prr(\tthetar) \EE_{\prior} [\indF(\btheta)|\Phir \tthetar] \left(\int_{\RR^{n-r}}   \pro(\tthetao)
d \tthetao \right)d \tthetar\\
=& \argmax_{v_r} \int_{\RR^r} [\ln \pibiasr(\tthetar; \bmv_r)]  \prr(\tthetar) \EE_{\prior} [\indF(\btheta)|\Phir \tthetar]d \tthetar\\
=& \argmax_{v_r} \int_{\RR^r} [\ln \pibiasr(\tthetar; \bmv_r) ] \prr(\tthetar) 
\left( \int_{\RR^{n-r}} \indF(\Phir \tthetar + \Phio \tthetao) \pro(\tthetao)d\tthetao \right)
d \tthetar\\
=& \argmax_{v_r} \int_{\RR^n} \indF(\Phir \tthetar + \Phio \tthetao)  [\ln \pibiasr(\tthetar; \bmv_r) ] \prior(\ttheta) 
d \ttheta\\
=& \argmax_{v_r}  \EE_{\prior} [\indF(\btheta)\ln \pibiasr(\tthetar; \bmv_r)].
\end{aligned} 
\end{equation}	
Again, we can use another $r$-dimensional biasing density $\pibias'$
with IS weights $w'(\tthetar) = \prr(\tthetar)/ \pibias'(\tthetar)$ to
evaluate the expectation in the last term of \cref{eq:CE-low}. Thus,
the optimal mean and covariance for the Gaussian proposal $\pibiasr$ in the low-dimensional subspace are
\begin{equation}\label{eq:mu-cov-low}
\mur^\star= \frac{\sum_{i=1}^N \indF(\btheta_i) w'(\tthetari) \tthetari }{\sum_{i=1}^N \indF(\btheta_i) w'(\tthetari) }, \; 
\Sigmar^\star= \frac{\sum_{i=1}^N \indF(\btheta_i) w'(\tthetari) (\tthetari-\mur)(\tthetari-\mur)^\top }{\sum_{i=1}^N \indF(\btheta_i) w'(\tthetari) },
\end{equation}
where $\tthetari\simiid \pibias'$,  $\tthetaoi \simiid \pro$ and $\btheta_i = \Phir \tthetari + \Phio \tthetaoi$.

\subsection{Multiple and adaptive importance sampling}\label{sec:MIS}
In the previous sections, we discussed using \LSIS as initial
biasing proposal $\pibias'=\pishift$ 
to estimate the mean and covariance of the optimal Gaussian
proposal. In this step, we already evaluate the parameter-to-event map
at several samples.
Naturally, the
question arises if we can reuse these samples in the IS, which
leads to multiple importance sampling (MIS) methods.  We
can also apply MIS in the CE estimation step, i.e., we reuse all previous
evaluations of $F$ in sampling to iteratively update the CE mean and covariance.
Repeated application of this procedure results in an
adaptive importance sampling method \cite{bugallo2017adaptive,paananen2021implicitly}, whose
adaptation to \cref{eq:prob-fail} is described next.

Multiple importance sampling (MIS) combines the independent samples
from different proposal distributions and builds a new IS estimator as
a weighted combination of different IS estimators
\cite{hesterberg1995weighted,veach1995optimally,owen2000safe}.  For
the problem we consider here, assume we have $J$ proposals as $\{\pi_j
\}_{j=1}^J$ and $N_j$ independent samples from each proposal
$\{\btheta_i^{(j)} \}_{i=1}^{N_j}\simiid \pi_j$, and we denote the
total number of samples as $N=\sum_{j=1}^J N_j$  and assume that the IS weight formula for each proposal is $w_j(\btheta) =
\prior(\btheta)/ \pi_j(\btheta)$. The MIS estimator
is defined as
\begin{equation}
\label{eq:est-MIS}
\pMIS = \frac{1}{N}\sum_{j=1}^{J} \sum_{i=1}^{N_j} \indF(\btheta_i^{(j)}) \wMIS (\btheta_i^{(j)}), 
\end{equation}
where weights $\wMIS$ are computed using one of the following
two strategies.
\paragraph{Standard MIS (s-MIS)}
This approach uses the same weights as IS estimator for each
individual proposal \cite{cappe2004population}. Thus, the MIS
estimator $\pMIS$ with standard proposal weights can be viewed as a weighted average of different IS estimators, with weights
proportional to $N_j$:
\begin{equation}
\label{eq:w-MIS-s}
\wMISs (\btheta_i^{(j)}) \coloneqq \frac{\prior(\btheta_i^{(j)})}{\pi_{j}(\btheta_i^{(j)})} = w_j(\btheta_i^{(j)}).
\end{equation}

\paragraph{Deterministic mixture MIS (dm-MIS)}
This approach uses weights defined as, \cite{owen2000safe},
\begin{equation}
\label{eq:w-MIS}
\wMISDM (\btheta_i^{(j)}) \coloneqq {\prior(\btheta_i^{(j)})}\bigg/{\sum\limits_{j'=1}^{J} \frac{N_{j'}}{N} \pi_{j'}(\btheta_i^{(j)})} 
= {N}\bigg/{\sum\limits_{j'=1}^{J} \frac{N_{j'}}{w_{j'}(\btheta_i^{(j)})} }.
\end{equation}
It can be shown that this dm-MIS estimator has
a variance that is smaller than
the variance of any weighting scheme, plus a term that goes to zero as
$\min\{N_j\}\to\infty$, \cite{veach1995optimally}.
One can also view $\pMIS$ with deterministic mixture weights as an IS
estimator using the biasing density $\piMIS$, which is a weighted
mixture distribution of $\{\pi_j \}_{j=1}^J$, i.e.,
\begin{equation}
\label{eq:pi-MIS}
\piMIS (\btheta) \coloneqq \sum\limits_{j'=1}^{J} \frac{N_{j'}}{N} \pi_{j'}(\btheta).
\end{equation}

In our adaptive importance sampling method, 
we reuse all evaluations of the parameter-to-event map
$\{\indF(\btheta_i^{(j)})\}_{i,j}$ together with the weights $\{
w_{j}(\btheta_i^{(j)})\}_{i,j}$ to compute
$\pMIS$. Moreover, we reuse all available sample
evaluations and weights
in each step of the CE method to update the Gaussian proposal.
Thus, the updated proposal $\pi_j$ depends on the previous
samples. The s-MIS estimator stays unbiased \cite{cappe2004population},
while the DM-MIS estimator might have a bias due to the loss of
independency between samples from different
proposals\cite{cornuet2012adaptive}. Although the dm-MIS estimator
cannot be proven to be unbiased, it generally outperforms other AIS methods and has smaller
errors in numerical experiments \cite{cornuet2012adaptive}. We also observe
this in our numerical tests, in which we compare results
obtained with s-MIS and dm-MIS weights.

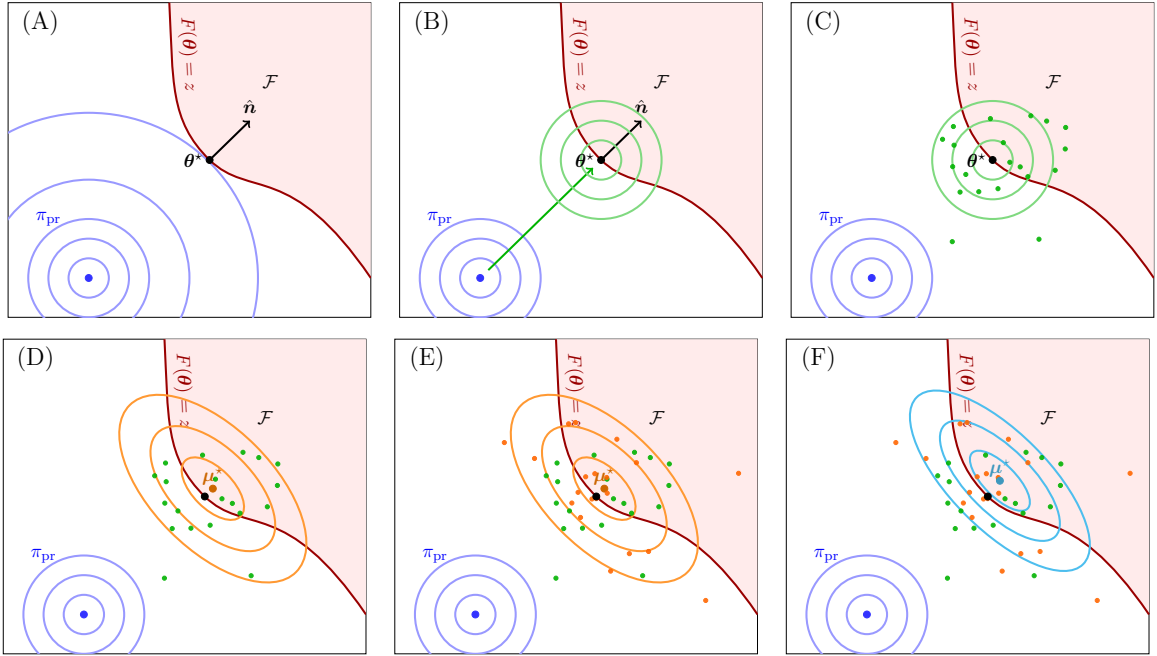
\begin{figure}[tb]
	\centering
		\begin{tikzpicture}[scale=0.65]
	\begin{axis}[compat=1.11, width=9cm, height=8cm, 
	xmin=-5,
	xmax=4,
	ymin=-4,
	ymax=4,
	tick style={draw=none},
	yticklabels={,,},
	xticklabels={,,},
	]
		\draw [blue!40,very thick] (-3,-3) circle (4.2);
		\draw [blue!40,very thick] (-3,-3) circle (2.5);
	\draw [blue!40,very thick] (-3,-3) circle (1.5);
	\draw [blue!40,very thick] (-3,-3) circle (1);
	\draw [blue!40,very thick] (-3,-3) circle (0.5);
	\node[blue] at (-4,-1.5) {$\prior$};
	\filldraw [fill=red!8,draw=none]
	(4,-3) .. controls (2,0) and (1,-1)
	.. (0,0).. controls (-1,1) and (-0.9,2) .. (-1,4) -- (4,4)  -- (4,-3);
	\draw[red!60!black,very thick]  (4,-3) .. controls (2,0) and (1,-1)
	.. (0,0).. controls (-1,1) and (-0.9,2) .. (-1,4) node[red!60!black,near end, above, sloped]
	{$F(\btheta)=z$};
	\node[] at (1.5,2) {$\FF$};
		\draw [black, very thick, ->] (0,0)	--
		(1,1) node[above]{$\hatn$}; 
	\addplot[mark=*] coordinates {(0,0)} ; 	
		\node[] at (-0.4,0){$\bthetastar$};
	\addplot[blue!80, mark=*] coordinates {(-3,-3)} ;
	\node[] at (-4.2,3.5){\Large{(A)}};
	\end{axis}
	\end{tikzpicture}
			\begin{tikzpicture}[scale=0.65]
	\begin{axis}[compat=1.11, width=9cm, height=8cm,
	xmin=-5,
	xmax=4,
	ymin=-4,
	ymax=4,
	tick style={draw=none},
	yticklabels={,,},
	xticklabels={,,},
	]
	\draw [blue!40,very thick] (-3,-3) circle (1.5);
	\draw [blue!40,very thick] (-3,-3) circle (1);
	\draw [blue!40,very thick] (-3,-3) circle (0.5);
	\node[blue] at (-4,-1.5) {$\prior$};
	\filldraw [fill=red!8,draw=none]
	(4,-3) .. controls (2,0) and (1,-1)
	.. (0,0).. controls (-1,1) and (-0.9,2) .. (-1,4) -- (4,4)  -- (4,-3);
	\draw[red!60!black,very thick]  (4,-3) .. controls (2,0) and (1,-1)
	.. (0,0).. controls (-1,1) and (-0.9,2) .. (-1,4) node[red!60!black,near end, above, sloped]
	{$F(\btheta)=z$};
	\node[] at (1.5,2) {$\FF$};
	\draw [black, very thick, ->] (0,0)	--
	(1,1) node[above]{$\hatn$}; 
	\draw [green!70!black, very thick, ->] (-2.8,-2.8)	--
	(-0.2,-0.2); 
	\draw[green!70!black!50, very thick] (0, 0) circle (0.5);
	\draw[green!70!black!50, very thick] (0, 0) circle (1);
	\draw[green!70!black!50, very thick] (0, 0) circle (1.5);
	\addplot[mark=*] coordinates {(0,0)} ; 	
	\node[] at (-0.4,0){$\bthetastar$};
	\addplot[blue!80, mark=*] coordinates {(-3,-3)} ;
		\node[] at (-4.2,3.5){\Large{(B)}};
	\end{axis}
	\end{tikzpicture}
			\begin{tikzpicture}[scale=0.65]
	\begin{axis}[compat=1.11, width=9cm, height=8cm,
	xmin=-5,
	xmax=4,
	ymin=-4,
	ymax=4,
	tick style={draw=none},
	yticklabels={,,},
	xticklabels={,,},
	]
	\draw [blue!40,very thick] (-3,-3) circle (1.5);
	\draw [blue!40,very thick] (-3,-3) circle (1);
	\draw [blue!40,very thick] (-3,-3) circle (0.5);
	\node[blue] at (-4,-1.5) {$\prior$};
	\filldraw [fill=red!8,draw=none]
	(4,-3) .. controls (2,0) and (1,-1)
	.. (0,0).. controls (-1,1) and (-0.9,2) .. (-1,4) -- (4,4)  -- (4,-3);
	\draw[red!60!black,very thick]  (4,-3) .. controls (2,0) and (1,-1)
	.. (0,0).. controls (-1,1) and (-0.9,2) .. (-1,4) node[red!60!black,near end, above, sloped]
	{$F(\btheta)=z$};
	\node[] at (1.5,2) {$\FF$};
	\draw[green!70!black!50, very thick] (0, 0) circle (0.5);
	\draw[green!70!black!50, very thick] (0, 0) circle (1);
	\draw[green!70!black!50, very thick] (0, 0) circle (1.5);
	\addplot [color=green!70!black, only marks, mark=*,mark size=1.2pt,
	opacity=0.9]
	table[x=x,y=y] {\data/sample.txt};
	\addplot[mark=*] coordinates {(0,0)} ; 	
	\node[] at (-0.4,0){$\bthetastar$};
	\addplot[blue!80, mark=*] coordinates {(-3,-3)} ;
		\node[] at (-4.2,3.5){\Large{(C)}};
	\end{axis}
	\end{tikzpicture}
			\begin{tikzpicture}[scale=0.65]
	\begin{axis}[compat=1.11, width=9cm, height=8cm,
	xmin=-5,
	xmax=4,
	ymin=-4,
	ymax=4,
	tick style={draw=none},
	yticklabels={,,},
	xticklabels={,,},
	]
	\draw [blue!40,very thick] (-3,-3) circle (1.5);
	\draw [blue!40,very thick] (-3,-3) circle (1);
	\draw [blue!40,very thick] (-3,-3) circle (0.5);
	\node[blue] at (-4,-1.5) {$\prior$};
	\filldraw [fill=red!8,draw=none]
	(4,-3) .. controls (2,0) and (1,-1)
	.. (0,0).. controls (-1,1) and (-0.9,2) .. (-1,4) -- (4,4)  -- (4,-3);
	\draw[red!60!black,very thick]  (4,-3) .. controls (2,0) and (1,-1)
	.. (0,0).. controls (-1,1) and (-0.9,2) .. (-1,4) node[red!60!black,near end, above, sloped]
	{$F(\btheta)=z$};
	\node[] at (1.5,2) {$\FF$};
	\addplot [color=green!70!black, only marks, mark=*,mark size=1.2pt,
	opacity=0.9]
	table[x=x,y=y] {\data/sample.txt};
	\addplot[orange!80!black, mark=*] coordinates {(0.2,0.2)}  node[yshift=0.25cm,xshift=0cm]{$\mustar$};
	\draw[orange!80, very thick, rotate around={45:(0.2,0.2)}] (0.2,0.2) ellipse (0.5 and 1);
	\draw[orange!80, very thick, rotate around={45:(0.2,0.2)}] (0.2,0.2) ellipse (1 and 2);
	\draw[orange!80, very thick, rotate around={45:(0.2,0.2)}] (0.2,0.2) ellipse (1.5 and 3);
	\addplot[mark=*] coordinates {(0,0)} ; 	
		\node[] at (-4.2,3.5){\Large{(D)}};
	\addplot[blue!80, mark=*] coordinates {(-3,-3)} ;
	\end{axis}
	\end{tikzpicture}
			\begin{tikzpicture}[scale=0.65]
	\begin{axis}[compat=1.11, width=9cm, height=8cm,
	xmin=-5,
	xmax=4,
	ymin=-4,
	ymax=4,
	tick style={draw=none},
	yticklabels={,,},
	xticklabels={,,},
	]
	\draw [blue!40,very thick] (-3,-3) circle (1.5);
	\draw [blue!40,very thick] (-3,-3) circle (1);
	\draw [blue!40,very thick] (-3,-3) circle (0.5);
	\node[blue] at (-4,-1.5) {$\prior$};
	\filldraw [fill=red!8,draw=none]
	(4,-3) .. controls (2,0) and (1,-1)
	.. (0,0).. controls (-1,1) and (-0.9,2) .. (-1,4) -- (4,4)  -- (4,-3);
	\draw[red!60!black,very thick]  (4,-3) .. controls (2,0) and (1,-1)
	.. (0,0).. controls (-1,1) and (-0.9,2) .. (-1,4) node[red!60!black,near end, above, sloped]
	{$F(\btheta)=z$};
	\node[] at (1.5,2) {$\FF$};
	\addplot [color=green!70!black, only marks, mark=*,mark size=1.2pt,
	opacity=0.9]
	table[x=x,y=y] {\data/sample.txt};
	\addplot[orange!80!black, mark=*] coordinates {(0.2,0.2)}  node[yshift=0.25cm,xshift=0cm]{$\mustar$};
	\draw[orange!80, very thick, rotate around={45:(0.2,0.2)}] (0.2,0.2) ellipse (0.5 and 1);
	\draw[orange!80, very thick, rotate around={45:(0.2,0.2)}] (0.2,0.2) ellipse (1 and 2);
	\draw[orange!80, very thick, rotate around={45:(0.2,0.2)}] (0.2,0.2) ellipse (1.5 and 3);
	\addplot [color=orange!80!red, only marks, mark=*,mark size=1.2pt,
	opacity=0.9]
	table[x=x,y=y] {\data/sample2.txt};
	\addplot[mark=*] coordinates {(0,0)} ; 	
	\addplot[blue!80, mark=*] coordinates {(-3,-3)} ;
		\node[] at (-4.2,3.5){\Large{(E)}};
	\end{axis}
	\end{tikzpicture}
			\begin{tikzpicture}[scale=0.65]
	\begin{axis}[compat=1.11, width=9cm, height=8cm,
	xmin=-5,
	xmax=4,
	ymin=-4,
	ymax=4,
	tick style={draw=none},
	yticklabels={,,},
	xticklabels={,,},
	]
	\draw [blue!40,very thick] (-3,-3) circle (1.5);
	\draw [blue!40,very thick] (-3,-3) circle (1);
	\draw [blue!40,very thick] (-3,-3) circle (0.5);
	\node[blue] at (-4,-1.5) {$\prior$};
	\filldraw [fill=red!8,draw=none]
	(4,-3) .. controls (2,0) and (1,-1)
	.. (0,0).. controls (-1,1) and (-0.9,2) .. (-1,4) -- (4,4)  -- (4,-3);
	\draw[red!60!black,very thick]  (4,-3) .. controls (2,0) and (1,-1)
	.. (0,0).. controls (-1,1) and (-0.9,2) .. (-1,4) node[red!60!black,near end, above, sloped]
	{$F(\btheta)=z$};
	\node[] at (1.5,2) {$\FF$};
	\addplot [color=green!70!black, only marks, mark=*,mark size=1.2pt,
	opacity=0.9]
	table[x=x,y=y] {\data/sample.txt};
	\addplot [color=orange!80!red, only marks, mark=*,mark size=1.2pt,
	opacity=0.9]
	table[x=x,y=y] {\data/sample2.txt};
	\addplot[mycyan!80!black, mark=*] coordinates {(0.3,0.4)}  node[yshift=0.25cm,xshift=0cm]{$\mustar$};
	\draw[mycyan, very thick, rotate around={45:(0.3,0.4)}] (0.3,0.4) ellipse (0.4 and 1);
	\draw[mycyan, very thick, rotate around={45:(0.3,0.4)}] (0.3,0.4) ellipse (0.8 and 2);
	\draw[mycyan, very thick, rotate around={45:(0.3,0.4)}] (0.3,0.4) ellipse (1.2 and 3);
	\addplot[mark=*] coordinates {(0,0)} ; 	
	\addplot[blue!80, mark=*] coordinates {(-3,-3)} ;
		\node[] at (-4.2,3.5){\Large{(F)}};
	\end{axis}
	\end{tikzpicture}
	\caption{Illustration of \RMIS: (A) Solve LDT optimization
          problem \cref{eq:LDT-opt} for $\bthetastar$, and construct
          low-dimensional subspace based on local Hessian information;
          (B) Adaptive importance sampling: start with \LSIS (green
          level sets); (C) AIS: Generate samples from \LSIS
          (green dots); (D) AIS: Use CE to
          find Gaussian mean and covariance $\mustar$ and $\Sigmastar$ (contour lines in orange);
          (E) Generate new samples from $\NN(\mustar, \Sigmastar)$
          (orange dots); (F) Use MIS and CE to update $\mustar$ and
          $\Sigmastar$, reusing previous samples (orange and green
          dots) and evaluations. }\label{fig:illustration}
\end{figure}

\begin{algorithm}
	\caption{LDT-based adaptive importance sampling (\RMIS)
	}\label{alg:LDT-IS}
	\begin{algorithmic}[1]
		\STATE\textbf{Input:} Function $F(\btheta)$, $\#$ of
                samples per CE step $\nCE$, 
                max CE
                iter.\ $J_{\max}$, threshold $\epsilon>0$.
		\STATE Compute LDT optimizer $\bthetastar$ by solving
                \cref{eq:LDT-opt} \label{alg:LDTopt}
		\STATE Choose subspace rank $r$ using $\epsilon$ as
                in \eqref{eq:epsilon-r} and compute
                orthonormal basis
                $\Phir$ as in \cref{sec:LDT-low} \label{alg:low-rank}
		\STATE Set level counter $J\gets 1$, total $\#$ of samples $N\gets 0$
		\STATE Set initial mean $ \mur^{(1)} =
                \Phir^\top\bthetastar$ and covariance $ \Sigmar^{(1)}
                = \bI_r $ %
		\WHILE{True}
		\STATE Draw $\nCE$ samples
                $\big\{\tthetari^{(J)}\big\}_{i=1}^{\nCE} \simiid\NN\big(
                \mur^{(J)}, \Sigmar^{(J)}\big)$ %
		\IF{s-MIS weights, or $J = 1$}
		\STATE Compute weights $w_i^{(J)} \gets
                \wGauss\big(\tthetari^{(J)};  \mur^{(J)},
                \Sigmar^{(J)}\big)$ \label{alg:w-MIS1}
		\ELSE
		\STATE Update weights:
$
		w_i^{(j)} \gets ({N+\nCE})\bigg/\bigg({\frac{N}{w_i^{(j)}} +
                  \frac{\nCE}{\wGauss\big(\tthetari^{(j)}; \mur^{(J)},
                    \Sigmar^{(J)} \big)}}\bigg)$, $1\leq j\leq J-1$ and
		compute weights 
		for new samples:
$
		w_i^{(J)} \gets ({N + \nCE})\bigg/{\sum_{j=1}^J
                  \frac{\nCE}{\wGauss\big(\tthetari^{(J)}; \mur^{(j)},
                    \Sigmar^{(j)}\big)}}
		$ 
		\label{alg:w-MIS}
		\ENDIF
		\STATE Draw $\nCE$ samples $\big\{\btheta_i^{(J)} \big\}_{i=1}^{\nCE}
                \simiid \prior$ and adjust in low-dim.\ subspace
                $\btheta_i^{(J)} \gets \btheta_i^{(J)}  -
                \Phir\Phir^\top \btheta_i^{(J)} + \Phir
                \tthetari^{(J)}$ \label{alg:theta-adjust}
		\STATE Evaluate $d_i^{(J)} \gets
                \indF(\btheta_i^{(J)})$ %
		\STATE Update $N \gets N+\nCE$ and compute
		IS estimator $\pRed\gets \frac{1}{N}\sum_{j=1}^J\sum_{i=1}^{\nCE} d_i^{(j)} w_i^{(j)}$
		\IF{%
			$J \geq J_{\max}$}
		\STATE Break
		\ENDIF
		\STATE Compute updated CE mean and covariance 
		\begin{equation*}
                  \small \hspace{-5ex}
		\mur^{(J+1)}\! \coloneqq \frac{\sum_{j=1}^J\sum_{i=1}^{\nCE}d_i^{(j)} w_i^{(j)} \tthetari^{(j) }}{\sum_{j=1}^J\sum_{i=1}^{\nCE} d_i^{(j)} w_i^{(j)} }, \;
		\Sigmar^{(J+1)} \!\coloneqq \frac{\sum_{j=1}^J\sum_{i=1}^{\nCE}d_i^{(j)} w_i^{(j)} (\tthetari^{(j)}-\mur^{(J+1)} )(\tthetari^{(j)}-\mur^{(J+1)} )^\top }{\sum_{j=1}^J\sum_{i=1}^{\nCE}d_i^{(j)} w_i^{(j)} }
		\end{equation*}
		\STATE $J \gets J+1$
		\ENDWHILE
		\STATE \textbf{Output:} $\pRed$
	\end{algorithmic}
\end{algorithm}

The steps of the proposed LDT-based adaptive sampling method (LAIS) are visualized in two dimensions in
\Cref{fig:illustration}. We start with a single proposal $\pishift$,
and repeatedly use MIS to reuse the previous samples and corresponding evaluations
to obtain an updated proposal using CE.
Combined with the LDT-informed dimension reduction, the initialization ($\pishift$ projected in the low-dimensional subspace), sampling and updates are conducted only in the low-dimensional subspace.
This results in an
adaptive multiple importance sampling scheme for rare event
probability estimation, which is summarized in
\Cref{alg:LDT-IS}. The first steps (\cref{alg:LDTopt,alg:low-rank}) in
this algorithm correspond to solving the LDT optimization problem \cref{eq:LDT-opt} and
finding a low-dimensional subspace as discussed in
\cref{sec:LDT-low}. Adaptive importance sampling in this
low-dimensional space is based on the CE method and multiple importance
sampling. 
The difference between \RMIS-s and \RMIS-dm lies in the computation of the MIS weights.
For \RMIS-s, the weights are computed using \cref{alg:w-MIS1}, i.e., the single proposal weights. For \RMIS-dm,
in the first iteration the IS weights are computed in
\cref{alg:w-MIS1}, and in iterations $J\ge 2$ they are updated in
\cref{alg:w-MIS}, which follows from \cref{eq:w-MIS}. This reuses the
previously computed weights $\{ w_i^{(j)}\}_{j=1}^{J-1}$. Only the IS
weights corresponding to $\wGauss$ (defined in \cref{eq:w-GIS}) need
to be recomputed. In \cref{alg:theta-adjust}, we combine subspace
samples from the CE-based parametric distribution with samples from
the prior in the orthogonal space. To avoid construction of the basis
$\Phio$ of the possibly high-dimensional space orthogonal to $\Phir$, we first
draw samples $\{\btheta_i^{(J)} \}_{i=1}^{\nCE}$ from the original
prior distribution $\prior$. Then we replace the component in the
low-dimensional subspace spanned by $\Phir$ with $\Phir
\tthetari^{(J)}$.

Note that \Cref{alg:LDT-IS} takes as input the number of CE
iterations, differently from most other CE methods that use a
stopping criterion to adaptively choose the number of iterations. For
example, a typical criterion 
is to use the quantile values to ensure that a good portion of samples
falls in $\FF$ \cite{de2005tutorial,papaioannou2019improved}.
Another criterion terminates the iterations when the coefficient of
variation of the estimator is smaller than a given threshold
\cite{uribe2021cross,papaioannou2019improved}. These stopping
criteria are usually implemented in the context of intermediate optimal
biasing densities, such as adaptively shrinking failure sets
\cite{de2005tutorial,papaioannou2019improved} or adaptively
smoothed indicator functions
\cite{uribe2021cross,papaioannou2019improved}.  Similar stopping
criteria can be used in \RMIS, but additional work is required to
fully justify their applicability.  Since the proposed \RMIS algorithm does not
use intermediate optimal biasing densities, using stopping rules such
as quantiles would require extra computations, which are not used
in other steps of \RMIS. As mentioned in
\cite{papaioannou2019improved}, the coefficient of variation of the
estimator might not be sufficiently robust when used as stopping
criterion. Thus the authors propose to use the CV of the ratio between
the indicator function and its smooth approximation instead. However,
this strategy is not applicable for our algorithm since we do not
use smoothed indicator functions as intermediate
densities. Another concern for adaptively choosing the number of CE
iterations based on the quantile or CV is that stopping rules
can introduce bias to the estimation, and it is hard to accurately
estimate the CV when samples may not be independent.
Thus, we use fixed numbers of CE steps in the algorithm to avoid additional
computations and bias, and to increase the robustness of the
method. Designing an adaptive stopping rule for \RMIS that
overcomes these challenges would be an interesting topic for future research.

In \cite{uribe2021cross}, a refinement step (Algorithm 2 of
\cite{uribe2021cross}) is added to \iCEred to further increase
accuracy. The basic idea is that the algorithm stops updating the
reference parameter, but continues sampling from the current biasing
density to evaluate the estimator, until reaching a desired
accuracy. A similar approach can also be used in \RMIS.

\section{Numerical experiments}
In this section, we compare the performance of the proposed \RMIS
variants (\RMIS-s using standard weights \cref{eq:w-MIS-s} and \RMIS-dm
using deterministic mixture weights \cref{eq:w-MIS} to compute the MIS estimator) with the \iCEred method \cite[Algorithm I]{uribe2021cross}. For
reference, we also include comparisons with \LSIS, which uses the
shift provided by the LDT optimizer and can thus be seen as a crude
version of \RMIS.
We use three example problems with increasing difficulties.
The first example is quadratic, and thus the LDT optimizer and the
eigenvectors and eigenvalues of $\HLDT$ are available
analytically. The other two examples involve the solution of
differential equations. For these examples, we use an
adjoint methods to compute gradients, and an interior-point method implemented in the
\texttt{fmincon} function in \textsc{MATLAB} to solve the
optimization problem \cref{eq:LDT-opt}.
The dominating eigenvectors of the matrix $\HLDT$ are computed using
\textsc{MATLAB}'s \texttt{eigs} function. This function only requires
the application of $\HLDT$ to vectors, which we approximate using finite
differences of gradients $\nabla F$. For \iCEred, the subspace is determined from
$\HFIS$, which requires the gradient of the logarithm of the smooth
approximation of $\indF(\btheta)$. We use the smooth approximation
$f(\btheta;s)$ defined in \cref{eq:smooth-cdf}.

\subsection{Quadratic parameter-to-event map $F$ from
  \cite{uribe2021cross}}
We first consider a
quadratic function $F$, which is the sum of a linear term and a
term that is quadratic in the first two components,
\begin{equation}\label{eq:LSF-Q}
F(\btheta)\coloneqq \frac{1}{\sqrt{n}}\sum_{k=1}^{n} \theta_k -\frac{\kappa}{4} (\theta_1-\theta_2)^2 , \; \btheta= [\theta_1, \ldots, \theta_n].
\end{equation}
Since $F$ is nonlinear only in the first two components, the
probability $\pF$ is independent of the dimension $n$.  The
remaining components result in a half-space and do not effect
the measure of the rare event set.
The first- and second-order derivatives of $F$ are:
\begin{align*}
\nabla F(\btheta) &= \left[\frac{\kappa}{2}(\theta_2-\theta_1)+\frac{1}{\sqrt{n}}, \frac{\kappa}{2}(\theta_1-\theta_2)+\frac{1}{\sqrt{n}}, \frac{1}{\sqrt{n}} \bm 1_{n-2}\right]^\top, \\
\nabla^2 F(\btheta) &=\left[  \begin{matrix}
-\frac{\kappa}{2} & \frac{\kappa}{2} &\\
\frac{\kappa}{2} & -\frac{\kappa}{2} & \\
& & \bzero_{n-2}
\end{matrix}\right],
\end{align*}
where $\bzero_{n-2}$ and $\bm 1_{n-2}$ denote vectors in $\mathbb
R^{n-2}$ consisting only of zeros and ones.
The LDT optimization \cref{eq:LDT-opt} is a quadratic programming
problem with analytic solution $\bthetastar = \frac{z}{\sqrt{n}}\bm 1_{n}$
and normal direction $\hatn = \frac{1}{\sqrt{n}}\bm 1_{n}$.  The
Hessian $\nabla^2 F(\btheta)$ has rank one, and the eigenvector
corresponding to its nonzero eigenvalue is $\bm y=[-\frac{\sqrt{2}}{2},
  \frac{\sqrt{2}}{2}, 0, \ldots, 0]^\top$.  Thus, the LDT-informed
subspace is spanned by $\{\hatn, \bm y\}$ and its dimension
is $r=2$ (independent of the choices of the threshold $\epsilon$). In all our experiments, the curvature parameter is chosen as
$\kappa=5$.
The coefficients of variation shown in this example are based on an average of
100 independent runs for each method. 

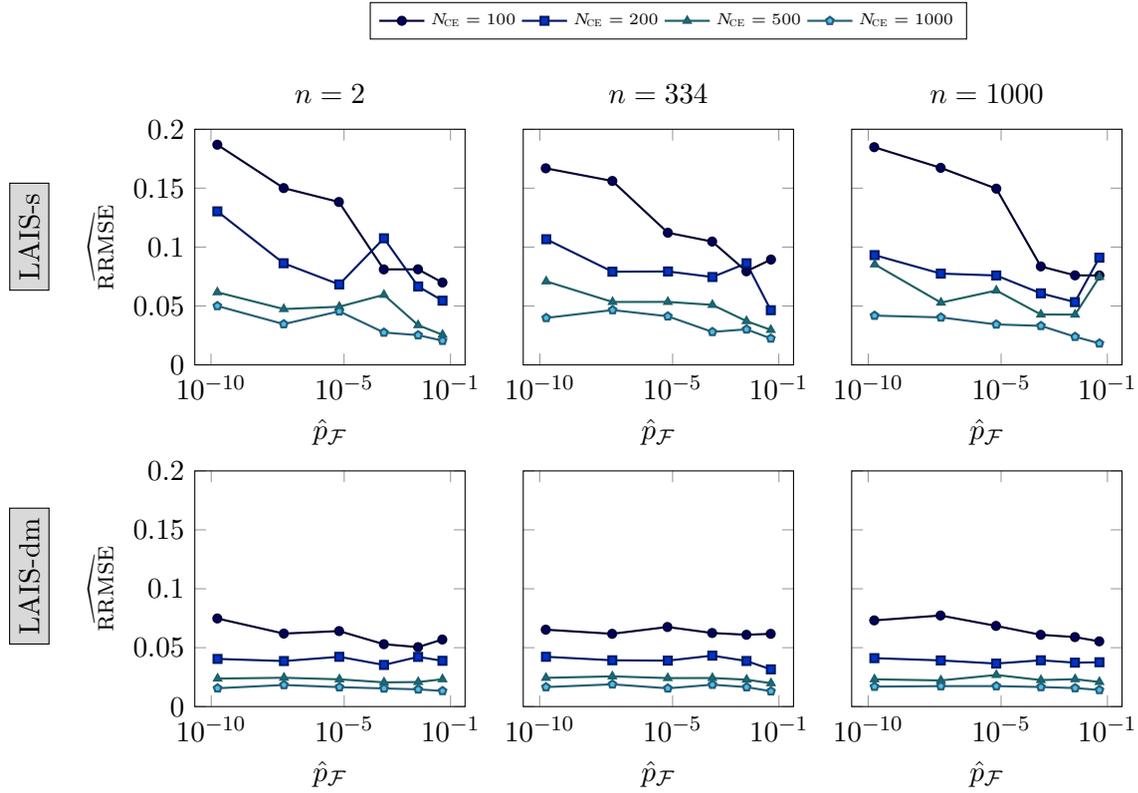
\begin{figure}[tb]
	\centering
	\begin{tikzpicture}[every mark/.append style={mark
              size=1.5pt}]
          \node[draw,rotate=90,fill=black!15!white] at (-2.2,1.7)
               {\RMIS-s};
          \node[draw,rotate=90,fill=black!15!white] at (-2.2,-2.8) {\RMIS-dm};

	\begin{groupplot}[group style = {rows=2, columns=3, horizontal
		sep = 22pt, vertical sep=40pt,}, width =  0.23\textwidth, height =
	0.2\textwidth, 
	xmode=log, 
	scale only axis,  
	cycle list name= my blue,
	compat = 1.3,
	ymin=0,
	ymax = 0.2,
	xlabel={$\hpF$},
	scaled y ticks=false,
	ytick={0,0.05,0.1,0.15,0.2},
	yticklabels={0,0.05,0.1,0.15,0.2},
	xtick={1e-10,1e-5,1e-1}
	]
	\nextgroupplot[
	ylabel={$\rrmse$},
	legend style = {font=\tiny,nodes=right, legend to name=grouplegend-fix-z-q},
	legend columns=4, 
	title = {$n=2$},
	]
	\addplot table[x =pF,y=rrmse] {\data/quad_z_RMIS_single_d2_nCE100.txt} ;
	\addlegendentry{$\nCE = 100$}
	\addplot table[x =pF,y=rrmse] {\data/quad_z_RMIS_single_d2_nCE200.txt} ;
	\addlegendentry{$\nCE = 200$}
	\addplot table[x =pF,y=rrmse] {\data/quad_z_RMIS_single_d2_nCE500.txt} ;
	\addlegendentry{$\nCE = 500$}
	\addplot table[x =pF,y=rrmse] {\data/quad_z_RMIS_single_d2_nCE1000.txt} ;
	\addlegendentry{$\nCE = 1000$}
	\nextgroupplot [yticklabels={,,},
	title = {$n=334$},]
	\addplot table[x =pF,y=rrmse] {\data/quad_z_RMIS_single_d334_nCE100.txt} ;
	\addplot table[x =pF,y=rrmse] {\data/quad_z_RMIS_single_d334_nCE200.txt} ;
	\addplot table[x =pF,y=rrmse] {\data/quad_z_RMIS_single_d334_nCE500.txt} ;
	\addplot table[x =pF,y=rrmse] {\data/quad_z_RMIS_single_d334_nCE1000.txt} ;
	\nextgroupplot [
	title = {$n=1000$}, yticklabels={,,}]
	\addplot table[x =pF,y=rrmse] {\data/quad_z_RMIS_single_d1000_nCE100.txt} ;
	\addplot table[x =pF,y=rrmse] {\data/quad_z_RMIS_single_d1000_nCE200.txt} ;
	\addplot table[x =pF,y=rrmse] {\data/quad_z_RMIS_single_d1000_nCE500.txt} ;
	\addplot table[x =pF,y=rrmse] {\data/quad_z_RMIS_single_d1000_nCE1000.txt} ;
		\nextgroupplot[
	ylabel={$\rrmse$},
	]
	\addplot table[x =pF,y=rrmse] {\data/quad_z_RMIS_DM_d2_nCE100.txt} ;
	\addplot table[x =pF,y=rrmse] {\data/quad_z_RMIS_DM_d2_nCE200.txt} ;
	\addplot table[x =pF,y=rrmse] {\data/quad_z_RMIS_DM_d2_nCE500.txt} ;
	\addplot table[x =pF,y=rrmse] {\data/quad_z_RMIS_DM_d2_nCE1000.txt} ;
	\nextgroupplot [yticklabels={,,},
	]
	\addplot table[x =pF,y=rrmse] {\data/quad_z_RMIS_DM_d334_nCE100.txt} ;
	\addplot table[x =pF,y=rrmse] {\data/quad_z_RMIS_DM_d334_nCE200.txt} ;
	\addplot table[x =pF,y=rrmse] {\data/quad_z_RMIS_DM_d334_nCE500.txt} ;
	\addplot table[x =pF,y=rrmse] {\data/quad_z_RMIS_DM_d334_nCE1000.txt} ;
	\nextgroupplot [yticklabels={,,},	%
	]
	\addplot table[x =pF,y=rrmse] {\data/quad_z_RMIS_DM_d1000_nCE100.txt} ;
	\addplot table[x =pF,y=rrmse] {\data/quad_z_RMIS_DM_d1000_nCE200.txt} ;
	\addplot table[x =pF,y=rrmse] {\data/quad_z_RMIS_DM_d1000_nCE500.txt} ;
	\addplot table[x =pF,y=rrmse] {\data/quad_z_RMIS_DM_d1000_nCE1000.txt} ;
	\end{groupplot}
	\node[black] at ($(group c1r1) + (4.5cm,3cm)$) {\pgfplotslegendfromname{grouplegend-fix-z-q}}; 
	\end{tikzpicture}
	\caption{Relative root-mean-square error of the rare event probability
		estimate for the quadratic example \cref{eq:LSF-Q} using
		different target probabilities. The number
		of CE updates in \Cref{alg:LDT-IS} is
                $J_{\max}=5$. The first row uses standard MIS weights,
                the second row deterministic mixture MIS weights.
	}\label{fig:quad-diff-z}
\end{figure}

In \Cref{fig:quad-diff-z}, we first study the performance of \RMIS-s and \RMIS-dm in dimensions $n\in\{2, 334, 1000\}$.
We use different values for the threshold, namely $z\in \{1,2,3,4,5,6\}$,
and compare the relative root-mean-square error (RRMSE, $\rrmse:=\sqrt{\EE[(\hpF-\pF)^2]}/\pF$)
after $J_{\max}=5$ CE updates. Note that the RRMSE requires knowledge of the
exact probability $\pF$, which is in general not available but can here
be computed accurately using numerical integration.
Since a different number of
samples $\nCE$ is used in each CE step, the overall number of samples
differs for different $\nCE$.
We observe a similar behavior for the different dimensions, which is a
consequence of using $r=2$ independently of the dimension $n$ and the
fact that the nonlinearity of $F$ is restricted to a two-dimensional subspace.
Most of the RRMSE for fixed $\nCE$ (\RMIS-s with $\nCE\geq200$ and \RMIS-dm with all choices of $\nCE$) are
insensitive to the target probability, which changes over 10
orders of magnitude. 
When increasing the number of samples per CE step, the
RRMSE decreases, consistent with the expected Monte Carlo convergence
rate, as $O(1/\sqrt{N})$ (note that since the total number of CE steps is
fixed, $O(1/\sqrt{N})=O(1/\sqrt{\nCE})$). In addition, \RMIS-dm is
insensitive to the dimension and target probability and has smaller absolute
errors. In particular, we do not observe bias due to the use of
deterministic mixture weights.

\begin{figure}[tb]
	\centering
	\begin{tikzpicture}[every mark/.append style={mark size=1.5pt}]
	\begin{groupplot}[group style = {rows=2, columns=3, horizontal
		sep = 30pt, vertical sep=40pt,}, width =  0.25\textwidth, height =
	0.2\textwidth, 
	xmode=log, scale only axis,  
	ymode =log,
	compat = 1.3,
	ymin=0,ymax=0.3,
	xmin = 800, xmax = 12000,
	cycle list name= quadratic,
	xlabel near ticks,
	ylabel near ticks,
	]
	\nextgroupplot[
	xlabel={$N$},
	ylabel={$\cv$},
	legend style = {font=\tiny,nodes=right, legend to name=grouplegend3},
	legend columns=3, 
	title = {$n=2$},
	]
	\addplot table[x =N,y=cv] {\data/quad_N_RMIS_single_d2_nCE200.txt} ;
	\addlegendentry{\RMIS-s, $\nCE = 200$}
	\addplot table[x =N,y=cv] {\data/quad_N_RMIS_single_d2_nCE500.txt} ;
	\addlegendentry{\RMIS-s, $\nCE = 500$}
	\addplot table[x =N,y=cv] {\data/quad_N_RMIS_single_d2_nCE1000.txt} ;
	\addlegendentry{\RMIS-s, $\nCE = 1000$}
	\addplot table[x =N,y=cv] {\data/quad_N_RMIS_DM_d2_nCE200.txt} ;
	\addlegendentry{\RMIS-dm, $\nCE = 200$}
	\addplot table[x =N,y=cv] {\data/quad_N_RMIS_DM_d2_nCE500.txt} ;
	\addlegendentry{\RMIS-dm, $\nCE = 500$}
	\addplot table[x =N,y=cv] {\data/quad_N_RMIS_DM_d2_nCE1000.txt} ;
	\addlegendentry{\RMIS-dm, $\nCE = 1000$}
	\addplot table[x =N,y=cv] {\data/quad_N_shift_d2.txt} ;
	\addlegendentry{\LSIS}
	\addplot table[x =N,y=cv] {\data/quad_CE_d2_delta1.5.txt} ;
	\addlegendentry{\iCEred}
	\nextgroupplot [xlabel={$N$}, title = {$n=334$},]
	\addplot table[x =N,y=cv] {\data/quad_N_RMIS_single_d334_nCE200.txt} ;
	\addplot table[x =N,y=cv] {\data/quad_N_RMIS_single_d334_nCE500.txt} ;
	\addplot table[x =N,y=cv] {\data/quad_N_RMIS_single_d334_nCE1000.txt} ;
	\addplot table[x =N,y=cv] {\data/quad_N_RMIS_DM_d334_nCE200.txt} ;
	\addplot table[x =N,y=cv] {\data/quad_N_RMIS_DM_d334_nCE500.txt} ;
	\addplot table[x =N,y=cv] {\data/quad_N_RMIS_DM_d334_nCE1000.txt} ;
	\addplot table[x =N,y=cv] {\data/quad_N_shift_d334.txt} ;
	\addplot table[x =N,y=cv] {\data/quad_CE_d334_delta1.5.txt} ;
	\nextgroupplot [xlabel={$N$}, title = {$n=1000$},]
	\addplot table[x =N,y=cv] {\data/quad_N_RMIS_single_d1000_nCE200.txt} ;
	\addplot table[x =N,y=cv] {\data/quad_N_RMIS_single_d1000_nCE500.txt} ;
	\addplot table[x =N,y=cv] {\data/quad_N_RMIS_single_d1000_nCE1000.txt} ;
	\addplot table[x =N,y=cv] {\data/quad_N_RMIS_DM_d1000_nCE200.txt} ;
	\addplot table[x =N,y=cv] {\data/quad_N_RMIS_DM_d1000_nCE500.txt} ;
	\addplot table[x =N,y=cv] {\data/quad_N_RMIS_DM_d1000_nCE1000.txt} ;
	\addplot table[x =N,y=cv] {\data/quad_N_shift_d1000.txt} ;
	\addplot table[x =N,y=cv] {\data/quad_CE_d1000_delta1.5.txt} ;
	\nextgroupplot[
	xlabel={$N$},
	ylabel={$\rrmse$},
	]
	\addplot table[x =N,y=rrmse] {\data/quad_N_RMIS_single_d2_nCE200.txt} ;
	\addplot table[x =N,y=rrmse] {\data/quad_N_RMIS_single_d2_nCE500.txt} ;
	\addplot table[x =N,y=rrmse] {\data/quad_N_RMIS_single_d2_nCE1000.txt} ;
	\addplot table[x =N,y=rrmse] {\data/quad_N_RMIS_DM_d2_nCE200.txt} ;
	\addplot table[x =N,y=rrmse] {\data/quad_N_RMIS_DM_d2_nCE500.txt} ;
	\addplot table[x =N,y=rrmse] {\data/quad_N_RMIS_DM_d2_nCE1000.txt} ;
	\addplot table[x =N,y=rrmse] {\data/quad_N_shift_d2.txt} ;
	\addplot table[x =N,y=rrmse] {\data/quad_CE_d2_delta1.5.txt} ;
	\nextgroupplot [xlabel={$N$}, 
	]
	\addplot table[x =N,y=rrmse] {\data/quad_N_RMIS_single_d334_nCE200.txt} ;
	\addplot table[x =N,y=rrmse] {\data/quad_N_RMIS_single_d334_nCE500.txt} ;
	\addplot table[x =N,y=rrmse] {\data/quad_N_RMIS_single_d334_nCE1000.txt} ;
	\addplot table[x =N,y=rrmse] {\data/quad_N_RMIS_DM_d334_nCE200.txt} ;
	\addplot table[x =N,y=rrmse] {\data/quad_N_RMIS_DM_d334_nCE500.txt} ;
	\addplot table[x =N,y=rrmse] {\data/quad_N_RMIS_DM_d334_nCE1000.txt} ;
	\addplot table[x =N,y=rrmse] {\data/quad_N_shift_d334.txt} ;
	\addplot table[x =N,y=rrmse] {\data/quad_CE_d334_delta1.5.txt} ;
	\nextgroupplot [xlabel={$N$}, 
	]
	\addplot table[x =N,y=rrmse] {\data/quad_N_RMIS_single_d1000_nCE200.txt} ;
	\addplot table[x =N,y=rrmse] {\data/quad_N_RMIS_single_d1000_nCE500.txt} ;
	\addplot table[x =N,y=rrmse] {\data/quad_N_RMIS_single_d1000_nCE1000.txt} ;
	\addplot table[x =N,y=rrmse] {\data/quad_N_RMIS_DM_d1000_nCE200.txt} ;
	\addplot table[x =N,y=rrmse] {\data/quad_N_RMIS_DM_d1000_nCE500.txt} ;
	\addplot table[x =N,y=rrmse] {\data/quad_N_RMIS_DM_d1000_nCE1000.txt} ;
	\addplot table[x =N,y=rrmse] {\data/quad_N_shift_d1000.txt} ;
	\addplot table[x =N,y=rrmse] {\data/quad_CE_d1000_delta1.5.txt} ;
	\end{groupplot}
	\node[black] at ($(group c1r1) + (4.5cm,3cm)$) {\pgfplotslegendfromname{grouplegend3}}; 
	\end{tikzpicture}
	\caption{Comparison of coefficient of variation and relative root-mean-square error versus overall
		number $N$ of sample evaluations for quadratic example
		\cref{eq:LSF-Q}. Shown is the behavior of different
		methods for fixed threshold $z=4$.
	}\label{fig:quad-diff-N}
\end{figure}
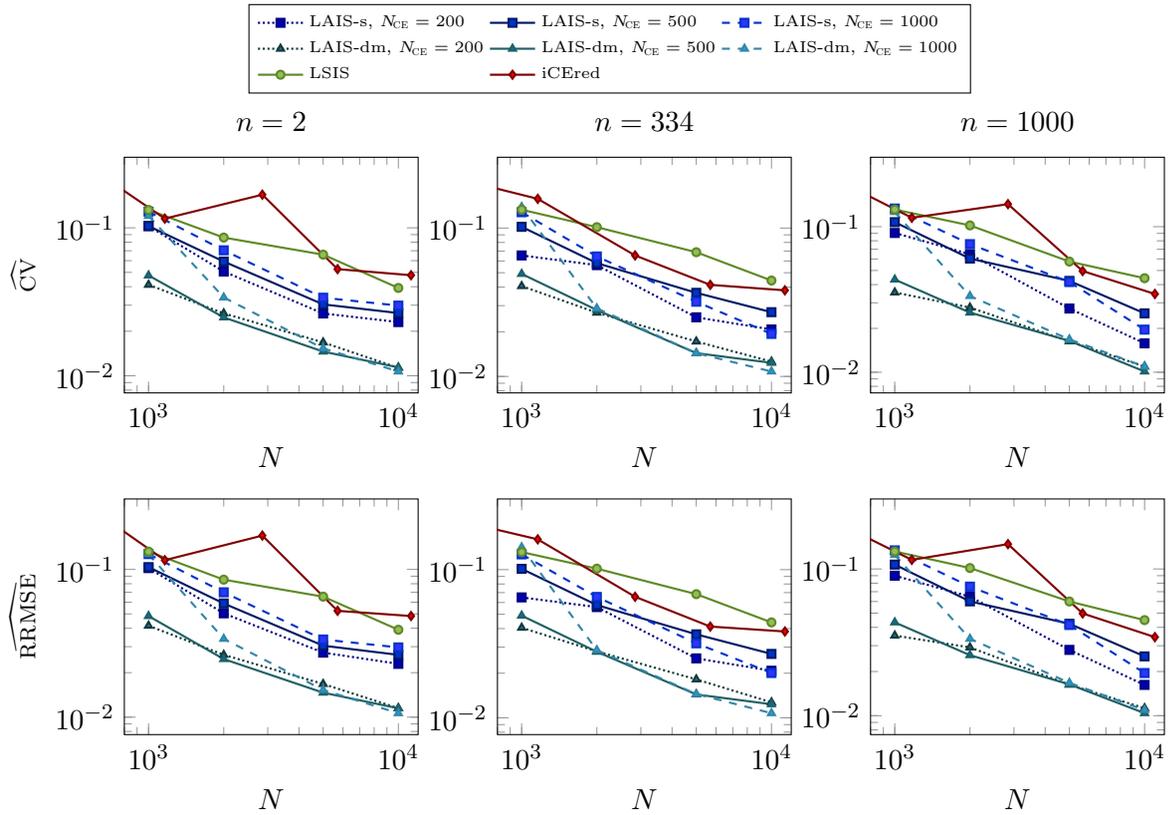

Next, in \Cref{fig:quad-diff-N}, we compare the performance of
different methods by plotting their estimated coefficient of variation (CV) and relative root-mean-square error (RRMSE) against the total number
$N$ of samples. This number is a measure for the overall computational
cost as we assume sample evaluations to be the dominating operation.
We compare both CV and RRMSE because \RMIS-dm might have a bias, which
would be visible in the RRMSE.
We fix $z=4$, resulting in a dimension-independent target
probability of $\pF\approx 6.62\times 10^{-6}$. For \iCEred,
we perform multiple experiments with different parameters and
decided to choose $\varepsilon=0.01$ and $\delta=1.5$ as in \cite{uribe2021cross}.
We
compare \RMIS (\RMIS-s and \RMIS-dm), \LSIS and \iCEred for different total numbers of
samples in different dimensions $n \in \{2, 334, 1000\}$. The behavior
of all methods is insensitive to the dimensions $n$ of the
random parameter, and RRMSE and CV are almost the same, demonstrating
that the bias in \RMIS-dm is neglectable.
Notice that \RMIS results in a much smaller CV and RRMSE compared with
\LSIS ($\cvshift \approx 4\times \cvRed, \cvshift \approx 1.5\times
\cvReds$, similar for $\rrmse$).
This is because the proposal used by  \LSIS is centered at
the boundary of the rare event set $\FF$, which in this problem is strictly convex in the
first two components. Thus, substantially more than half of the
samples fall outside the rare event set, where $\indF$ evaluates to zero.
Since \RMIS adjusts the mean and covariance of the Gaussian proposal,
the result is better tailored to the rare event set.
For the same number of total samples, the CV and RRMSE for \iCEred are
similar to \LSIS.
Hence, \RMIS-dm and \RMIS-s outperform \iCEred by factors of about 4
and 1.5 in the errors.
This is likely since \RMIS starts from a good proposal and reuses samples
while adaptively refining the mean and covariance. Thus the \RMIS
estimators uses the total $N$ evaluations while the \iCEred estimator
only uses the last $\nCE$ evaluations.

\begin{figure}[tb]
	\centering
	\begin{tikzpicture}[every mark/.append style={mark size=1.5pt}]
	\begin{groupplot}[group style = {rows=2, columns=3, horizontal
		sep = 30pt, vertical sep=40pt,}, width =  0.25\textwidth, height =
	0.2\textwidth, 
	xmode=log, scale only axis,  
	ymode =log,
	compat = 1.3,
	cycle list name= quadratic2,
	xlabel near ticks,
	ylabel near ticks,
	]
	\nextgroupplot[
	xlabel={$\hpF$},
	ylabel={$N$},
	legend style = {font=\tiny,nodes=right, legend to name=grouplegendcost},
	legend columns=4, 
	title = {$n=2$},
	]
	\addplot table[x =pF,y=N] {\data/quad_cost_RMIS_single_d2_delta1.5.txt} ;
	\addlegendentry{\RMIS-s}
	\addplot table[x =pF,y=N] {\data/quad_cost_RMIS_DM_d2_delta1.5.txt} ;
	\addlegendentry{\RMIS-dm}
	\addplot table[x =pF,y=N] {\data/quad_cost_LSIS_d2.txt} ;
	\addlegendentry{\LSIS}
	\addplot table[x =pF,y=N] {\data/quad_cost_CE_d2_delta1.5.txt} ;
	\addlegendentry{\iCEred}
	\nextgroupplot [xlabel={$\hpF$}, title = {$n=334$},]
	\addplot table[x =pF,y=N] {\data/quad_cost_RMIS_single_d334_delta1.5.txt} ;
	\addplot table[x =pF,y=N] {\data/quad_cost_RMIS_DM_d334_delta1.5.txt} ;
	\addplot table[x =pF,y=N] {\data/quad_cost_LSIS_d334.txt} ;
	\addplot table[x =pF,y=N] {\data/quad_cost_CE_d334_delta1.5.txt} ;
	\nextgroupplot [xlabel={$\hpF$}, title = {$n=1000$},]
	\addplot table[x =pF,y=N] {\data/quad_cost_RMIS_single_d1000_delta1.5.txt} ;
	\addplot table[x =pF,y=N] {\data/quad_cost_RMIS_DM_d1000_delta1.5.txt} ;
	\addplot table[x =pF,y=N] {\data/quad_cost_LSIS_d1000.txt} ;
	\addplot table[x =pF,y=N] {\data/quad_cost_CE_d1000_delta1.5.txt} ;
	\end{groupplot}
	\node[black] at ($(group c1r1) + (4.5cm,2.7cm)$) {\pgfplotslegendfromname{grouplegendcost}}; 
	\end{tikzpicture}
	\caption{Comparison of number of totally required samples $N$
          for different methods to obtain an RRMSE in the range $[4.5\times
            10^{-2}, 5\times 10^{-2}]$ using different target
          probabilities $\hpF$ and problem dimensions $n$.}\label{fig:quad-cost}
\end{figure}
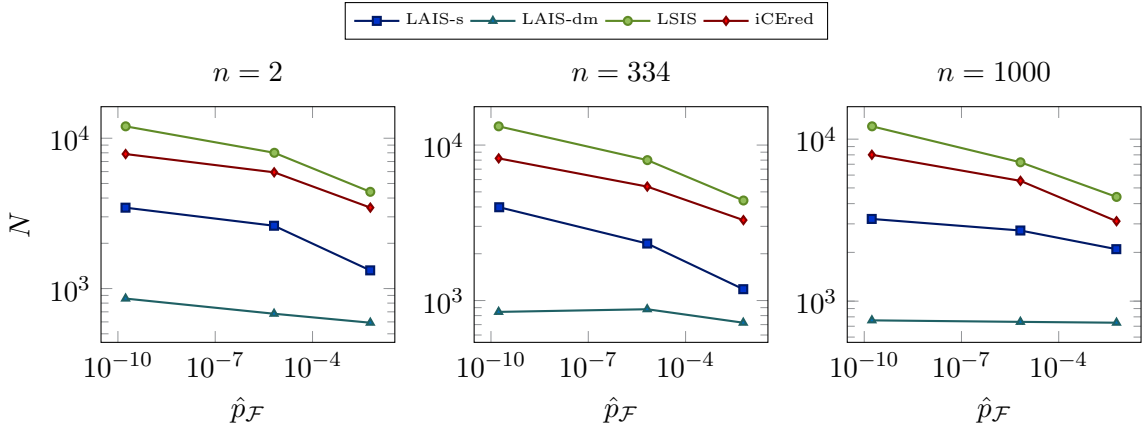

In \Cref{fig:quad-cost}, we compare the computational cost (total
number of samples $N$) for different methods to achieve a fixed RRMSE.
The figure is generated by varying different
$\nCE$ and the corresponding total number of iterations, using the
same parameter as for the experiments in
\Cref{fig:quad-diff-N}. 
The
computational costs of \LSIS and \iCEred are similar, and both methods
are outperformed by the \RMIS methods.
All methods depend weakly on the extremeness of the events, i.e., the required
sample size increases as the target probability decreases.
However, the required sample size for \RMIS-s and \RMIS-dm is more
robust to extremeness. For \RMIS-dm, the total number of
required samples remains mostly constant, demonstrating that \RMIS
becomes even more advantageous for rarer events.

\begin{figure}[tb]
	\centering
	\begin{tikzpicture}[scale=0.9]
	\begin{axis}[
	compat=1.11, 
	width=9cm, height=3.6cm, 
	xmin=1,
	xmax=7,
	ymin=-0.5,
	ymax=0.5,
	axis equal,
	xlabel = {$\hatn$},
	ylabel = {$\bm y$},
	title={$\kappa = 5$},
	]
	\filldraw [\bluetwo] (2.52,0) circle (2pt);
	\draw[\bluetwo, very thick] (2.52,0) ellipse (0.39 and 0.26); %
	\draw[\bluetwo, very thick] (2.52,0) ellipse (0.59 and 0.40); %
	\filldraw [\bluefour] (4.33,0) circle (2pt);
	\draw[\bluefour, very thick] (4.33,0) ellipse (0.26 and 0.21); %
	\draw[\bluefour, very thick] (4.33,0) ellipse (0.39 and 0.32); %
	\filldraw [\bluesix] (6.23,0) circle (2pt);
	\draw[\bluesix, very thick] (6.23,0) ellipse (0.19 and 0.18); %
	\draw[\bluesix, very thick] (6.23,0) ellipse (0.29 and 0.27); %
	\addplot[\bluetwo,very thick, densely dashed, domain=-1:1,samples=30]({5*x^2/2+2},{x});
	\addplot[\bluefour,very thick, densely dashed, domain=-1:1,samples=30]({5*x^2/2+4},{x});
	\addplot[\bluesix,very thick, densely dashed, domain=-1:1,samples=30]({5*x^2/2+6},{x});
	\node[rectangle, fill=gray!30!white] at (6.23,0.6){$z=6$};
	\node[rectangle, fill=gray!30!white] at (4.33,0.6){$z=4$};
	\node[rectangle, fill=gray!30!white] at (2.52,0.6){$z=2$};
        \end{axis}
	\end{tikzpicture}
		\begin{tikzpicture}[scale=0.9]
	\begin{axis}[
	compat=1.11, 
	width=9cm, height=7.5cm, 
	xmin=1,
	xmax=7,
	ymin=-2.5,
	ymax=2.5,
	axis equal,
	xlabel = {$\hatn$},
	ylabel = {$\bm y$},
	title={$\kappa = -0.1$},
	extra y ticks={-1.5, -0.5, 0.5,1.5},
	]
	\filldraw [\redtwo] (2.32,0) circle (2pt);
	\draw[\redtwo, very thick] (2.32,0) ellipse (0.35 and 1.14); %
	\draw[\redtwo, very thick] (2.32,0) ellipse (0.55 and 1.76); %
	\filldraw [\redfour] (4.15,0) circle (2pt);
	\draw[\redfour, very thick] (4.15,0) ellipse (0.25 and 1.30); %
	\draw[\redfour, very thick] (4.15,0) ellipse (0.38 and 2.00); %
	\filldraw [\redsix] (6.04,0) circle (2pt);
	\draw[\redsix, very thick] (6.04,0) ellipse (0.23 and 1.56); %
	\draw[\redsix, very thick] (6.04,0) ellipse (0.35 and 2.40); %
	\addplot[\redtwo,very thick, densely dashed, domain=-2.5:2.5,samples=30]({-0.1*x^2/2+2},{x});
	\addplot[\redfour,very thick, densely dashed, domain=-2.5:2.5,samples=30]({-0.1*x^2/2+4},{x});
	\addplot[\redsix,very thick, densely dashed, domain=-2.5:2.5,samples=30]({-0.1*x^2/2+6},{x});
	\node[rectangle, fill=gray!30!white] at (2.32,2.28){$z=2$};
	\node[rectangle, fill=gray!30!white] at (4.15,2.28){$z=4$};
	\node[rectangle, fill=gray!30!white] at (6.04,2.28){$z=6$};
	\end{axis}
	\end{tikzpicture}
	\caption{Comparison of mean and confidence regions of optimal
          Gaussian biasing distributions in the two-dimensional
          reduced subspace for the quadratic problem \cref{eq:LSF-Q}
          and for different $z$. The top figure is for $\kappa=5$ and
          the bottom one for $\kappa=-0.1$. The $x$-axis corresponds
          to the normal direction $\hatn$, and the $y$-axis to
          direction $\bm y$, the only non-zero curvature
          direction. The dots are the optimal means $\mustar$, and the
          ellipses show the $40\% $ and $70\%$ confidence regions.
          The dashed lines are the boundaries of the rare event sets,
          which are to the right to the boundary.
	}\label{fig:decay}
\end{figure}
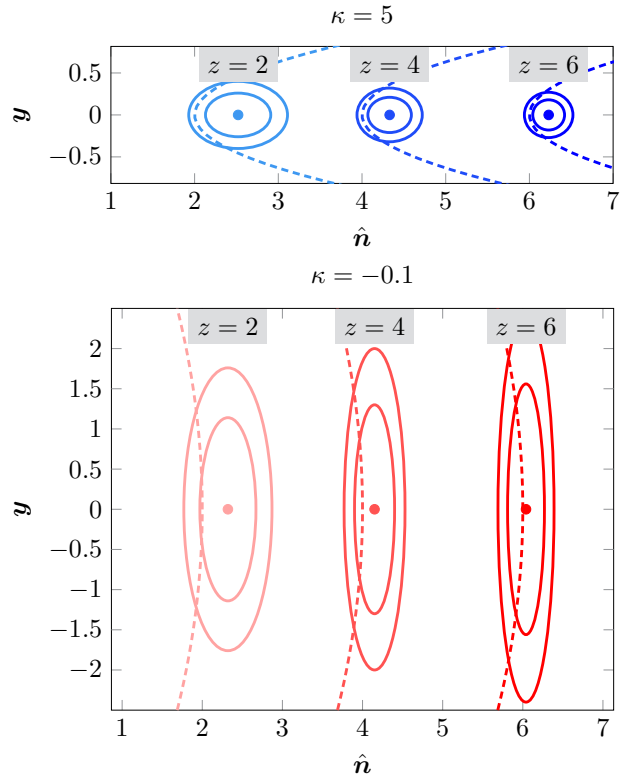

A possible concern when using tailored Gaussians as biasing densities is a
possible degeneration for large $z$, since, as predicted by LDT, we
expect that the optimal variance in the direction $\hatn$ decays
faster than in the orthogonal directions as $z$ increases. To study
if this is a problem in practice, we calculate the means and covariances of optimal Gaussian
distributions for $z\in\{
2,4, 6\}$ through numerical integration of their analytic
form. In \Cref{fig:decay}, we show the resulting $40\%$ and $70\%$
confidence regions in the reduced subspace spanned by normal and tangential
directions $\hatn$ and $\bm y$, respectively.
For $\kappa=5$ (top of \Cref{fig:decay}), the confidence region of the optimal
Gaussian distributions is reduced in all directions as
$z\to\infty$. As expected, it decays slightly faster in the normal
direction. However, we
do not observe a degeneration as the ratio only varies moderately over a wide range of
probabilities\footnote{For $z=6$, the probability $\pF$ is around
$10^{-10}$.} The results for $\kappa=-0.1$ are shown at the bottom of
\Cref{fig:decay}. Here, for growing $z$, the confidence region increases in the
$\bm y$-direction while it decreases in the normal direction
$\hatn$.
The optimal Gaussian becomes ``slim", but the ratio is still reasonable
and unlikely to lead to numerical problems. This suggests that
degeneration of approximating Gaussians and potentially resulting numerical
instabilities does not seem to be a substantial problem, at least for
this example.
If one encounters fast degeneracy of the optimal
Gaussian density in the normal direction, then special treatments in that
direction may be required. For example, one could use an exponential
density function
 in the normal direction. However, in that case, an explicit formula
 similar to \cref{eq:mu-cov} would not be available since the parametrized
 distribution would not be Gaussian; see also the discussion in 
 \cref{sec:discussion}.

Finally, due to the simplicity of \cref{eq:LSF-Q}, analytic expressions of the LDT optimizer
and the LDT-informed subspace are available.  In general, solving the
LDT optimization problem as well as finding the LDT-informed subspace
require numerical computations. These increases the computational cost of
\RMIS. Also \iCEred requires additional evaluations of gradients of
$F$ to find the low-rank subspaces. When the evaluation of $F$ is complicated,
e.g., when it involves the solution of a differential equation, the
evaluation of $F$ and its gradients typically dominate the overall
computational cost. The next two problems are examples where this is
the case.

\subsection{1D diffusion problem from \cite{wagner2020multilevel}}
Next, we consider a 1D diffusion problem defined on the domain
$D=[0,1]$. The stochastic differential equation
is
\begin{equation}
\label{eq:PDE-1D}
\begin{aligned}
-\frac{\partial}{\partial x} \left( a(x,\omega)\frac{\partial}{\partial x} v(x,\omega)\right)  = 1, &&  x\in[0,1], \\
v(0,\omega) = 0, &&\\
v'(1,\omega) = 0,
\end{aligned}
\end{equation}
for almost every $\omega\in\Omega$.  The coefficient function $a(x,
\omega)$ is a log-normal random field with mean $\EE[a(x,\cdot)]=1$
and $\VV[a(x,\cdot)] = 0.01$. Assuming that $a(x,\omega)= \exp(Z(x,\omega))$,
then $Z(x,\omega)$ is a Gaussian random field with mean
$\mu_Z=\ln(\EE(a,\cdot))-\sigma_Z^2/2$ and variance
$\sigma_Z^2=\ln((\VV[a(x,\cdot)] +\EE[a(x,\cdot)]^2
)/\EE[a(x,\cdot)]^2)$. The covariance operator is defined as
$c(x,y)=\sigma_Z^2\exp(-|x-y|/l)$ with correlation length
$l=0.01$. The truncated KL expansion of
$Z(x,\omega)$ is
\begin{equation}
\label{eq:KL-Z}
Z(x,\omega) = \mu_Z + \sum_{m=1}^n \sqrt{\lambda_m} e_m(x) \theta_m(\omega),
\end{equation}
where $(\lambda_m, e_m)$ are eigenpairs of $c(x,y)$,
and we use the dimension $n=150$. The random parameter $\btheta\in \RR^{n}$ and
$\btheta\sim \NN(\bzero, \bI_n)$.  The eigenpairs of $c(x,y)$ are
obtained using the Nystrom method \cite{press2007numerical} with one
Gauss-Legendre point in each element (512 equidistant elements in
total).

The event we are interested in is the value of the state at the
endpoint $x=1$, and we use the
threshold $z=0.535$, i.e., we are interested in the
probability that
\begin{equation}
F(\btheta)\coloneqq v(1,\omega) \ge 0.535.
\end{equation}
The corresponding rare event probability is $\pF\approx1.63\times10^{-4}$ (computed using \LSIS with $5\times 10^6$ samples).
We use a finite element method with 512 elements and linear basis
functions to discretize \cref{eq:PDE-1D}. The
adjoint method is used to obtain the gradient $\nabla F(\btheta)$.

\begin{figure}[tb]
	\centering
	\begin{tikzpicture}[every mark/.append style={mark size=1.5pt}]
	\begin{groupplot}[group style = {rows=1, columns=2, horizontal
		sep = 50pt, vertical sep=40pt,}, width =  0.4\textwidth, height =
	0.35\textwidth, 
	xmode=log, scale only axis,  
	ymode =log,
	compat = 1.3,
	ymin=0,ymax=0.15,
	xmin = 800, xmax = 35000,
	cycle list name= diffusion,
	xlabel near ticks,
	ylabel near ticks,
	ylabel shift = -10 pt,
	]
	\nextgroupplot[
	xlabel={$\NF$: number of evaluations of $F$},
	ylabel={$\cv$},
	legend style = {font=\tiny,nodes=right, legend to name=grouplegenddif},
	legend columns=3, 
	]
	\addplot table[x =N,y=cv] {\data/diffus_RMIS_single_nCE200_r3.txt} ;
	\addlegendentry{\RMIS-s, $\nCE=200, r=3$}
	\addplot table[x =N,y=cv] {\data/diffus_RMIS_single_nCE500_r3.txt} ;
	\addlegendentry{\RMIS-s, $\nCE=500, r=3$}
	\addplot table[x =N,y=cv] {\data/diffus_RMIS_single_nCE500_r5.txt} ;
	\addlegendentry{\RMIS-s, $\nCE=500, r=5$}
	\addplot table[x =N,y=cv] {\data/diffus_RMIS_DM_nCE200_r3.txt} ;
	\addlegendentry{\RMIS-dm, $\nCE=200, r=3$}
	\addplot table[x =N,y=cv] {\data/diffus_RMIS_DM_nCE500_r3.txt} ;
	\addlegendentry{\RMIS-dm, $\nCE=500, r=3$}
	\addplot table[x =N,y=cv] {\data/diffus_RMIS_DM_nCE500_r5.txt} ;
	\addlegendentry{\RMIS-dm, $\nCE=500, r=5$}
	\addplot table[x =N,y=cv] {\data/diffus_shift.txt} ;
	\addlegendentry{\LSIS}
	\addplot table[x =N,y=cv] {\data/diffus_CE_delta1.5_eps0.05.txt} ;
	\addlegendentry{\iCEred, $\delta=1.5, \varepsilon=0.05$}
	\addplot table[x =N,y=cv] {\data/diffus_CE_delta0.8_eps0.05.txt} ;
	\addlegendentry{\iCEred, $\delta=0.8, \varepsilon=0.05$}
	\nextgroupplot [
	xlabel={$\NF$: number of evaluations of $F$},
	ylabel={$\rrmse$},
	]
	\addplot table[x =N,y=rrmse] {\data/diffus_RMIS_single_nCE200_r3.txt} ;
	\addplot table[x =N,y=rrmse] {\data/diffus_RMIS_single_nCE500_r3.txt} ;
	\addplot table[x =N,y=rrmse] {\data/diffus_RMIS_single_nCE500_r5.txt} ;
	\addplot table[x =N,y=rrmse] {\data/diffus_RMIS_DM_nCE200_r3.txt} ;
	\addplot table[x =N,y=rrmse] {\data/diffus_RMIS_DM_nCE500_r3.txt} ;
	\addplot table[x =N,y=rrmse] {\data/diffus_RMIS_DM_nCE500_r5.txt} ;
	\addplot table[x =N,y=rrmse] {\data/diffus_shift.txt} ;
	\addplot table[x =N,y=rrmse] {\data/diffus_CE_delta1.5_eps0.05.txt} ;
	\addplot table[x =N,y=rrmse] {\data/diffus_CE_delta0.8_eps0.05.txt} ;
	\end{groupplot}
	\node[black] at ($(group c1r1) + (3.5cm,4cm)$) {\pgfplotslegendfromname{grouplegenddif}}; 
	\end{tikzpicture}
	\caption{Comparison of coefficient of variation and relative root-mean-square error plotted
		against number $N_F$ of evaluations of $F$ for 1D
		diffusion problem. Shown are results obtained with \RMIS,
		\LSIS and \iCEred with various parameter
		choices.
	}\label{fig:diff-N}
\end{figure}
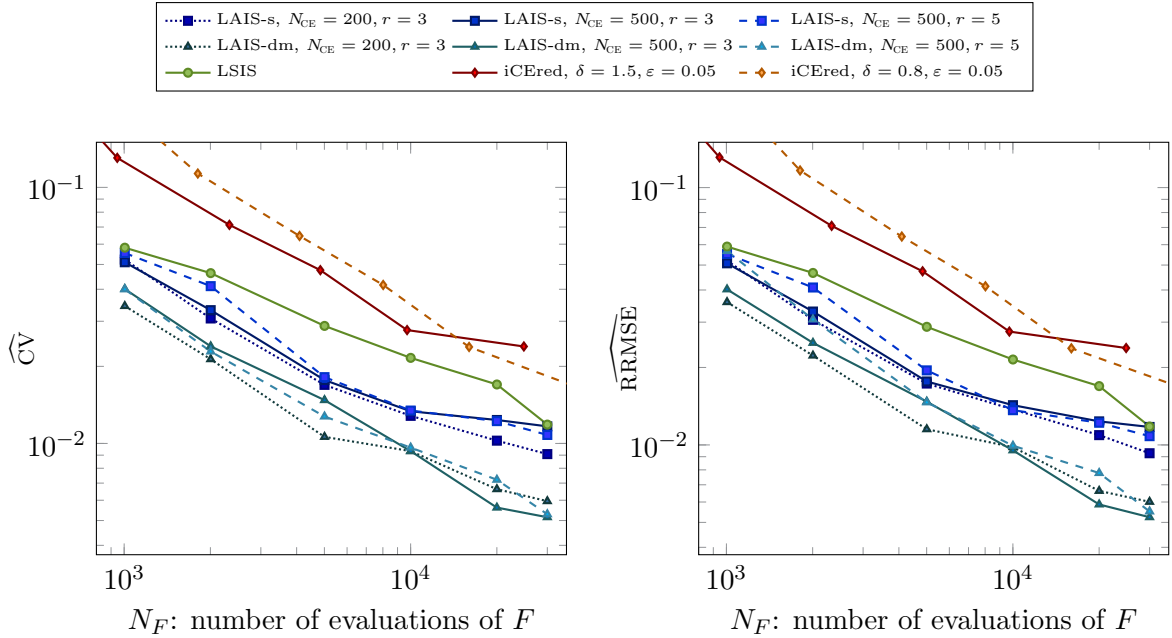

In \Cref{fig:diff-N}, we compare \RMIS (\RMIS-s and \RMIS-dm), \iCEred and \LSIS  with
respect to the required total number of evaluations of $F$
(denoted as $\NF$). The results are from an average of 100 independent
runs of each method. The rank $r$ for \RMIS is chosen based on the
absolute values of eigenvalues of $\HLDT$.
For the different algorithms, we estimate the RRMSE and CV, which are
again similar. The only difference occurs for \RMIS-dm with
$\nCE=500, r=5$ when $\NF\leq 2\times 10^3$, which is mainly due to the
dependency between samples being stronger for the first iterations
($J\leq 4$) when the dimension of reference parameter $O(r^2)$ is
large and the sample size $\nCE$ is small. These discrepancies
disappear after several iterations. Thus, the potential bias in \RMIS-dm is
controllable.

In general, the LDT-informed methods \LSIS and \RMIS
outperform \iCEred. 
\LSIS and \iCEred perform more
similarly as the total number of samples increases, but \LSIS still outperforms \iCEred by a factor of 1.5 in terms of errors. 
The RRMSE and CV of \RMIS-dm are more than 3 times smaller than \iCEred (2.5 times smaller than \LSIS), those of \RMIS-s are more than 2 times smaller than \iCEred (1.5 times smaller than \LSIS).

To study the robustness of the algorithms, we
vary the number of samples per CE step $\nCE=\{200, 500\}$ and the
rank $r=\{3, 5\}$ used for the subspace in \RMIS. The latter ranks
correspond to
the thresholds $\epsilon=0.075$ for $r=3$ and
$\epsilon=0.065$ for $r=5$ in \eqref{eq:epsilon-r}.
In \iCEred, we use different
$\delta=\{0.8, 1.5\}$, and $\varepsilon=0.05$.
We choose $\nCE$ in \iCEred such that we obtain different total
numbers of evaluations of $F$.
The average number of levels for \iCEred is 4, and the final rank depends on $\nCE$, with an average: $r=6$ for $\nCE=200$ and $r=12$ for $\nCE=500$, which is slightly larger than the ones we use for \RMIS.
\RMIS
with different ranks of the subspace and different $\nCE$ per CE steps perform similarly,
demonstrating that the algorithm is rather insensitive to these
choices.
We find the performance of \iCEred depends on
$\delta$, which controls the decay of the smoothing parameter $s$ from
level to level, and at what point the method terminates.

Note that in \Cref{fig:diff-N}, we only report $\NF$, the number of
evaluations of $F$. For \RMIS and \LSIS, this number is $\NF=N+5$, where $N$
is the total number of samples, and the extra five evaluations come
from solving the constrained optimization.
For \iCEred, this number
is the same as the total number of samples $\NF=N$.  However, there is
a large difference in the number of gradient
evaluations (we denote this number by $\NdF$) in
\RMIS, \LSIS and \iCEred. For \RMIS and \LSIS, the optimization step
requires 4 gradient computations, while the eigenvalue estimation
requires 50-60 gradients, thus the total number of gradient
computations is $\NdF\leq70$, largely independent of
of $r$ and $\nCE$. The \iCEred method does not require solution of an
optimization problem, but constructs the matrix
\cref{eq:H-FIS} for every level, which requires evaluation of $\nabla F$
for each evaluation of $F$. Thus, the number of gradient evaluations
for \iCEred is $\NdF=N$. Since we use an adjoint method to compute
gradients, one gradient evaluation amounts to solving an additional
boundary value problem (the adjoint equation). Thus, the total cost
(ODE/PDE solves) for \iCEred is $\NF+\NdF \approx2N$, while the number
for \RMIS and \LSIS remains approximately $N$.

\subsection{Tsunami problem}
Finally, we consider a more realistic problem, where we estimate the
probability of the extreme tsunami wave height on shore. This example
is adapted from Section 5 in \cite{tong2021extreme}.
To model tsunami waves, we use the one-dimensional shallow water
equations \cite{leveque2008high} defined on a domain $\mathcal{D}=[a,b]$
for times $t\in[0,T_F]$. The domain represents a slice through the sea,
that includes the shallow part near the shore and the part
where the ocean floor elevation can change unpredictably during an earthquake.
We denote the horizontal fluid velocity as $u(x,t)$ and the height of
water above the ocean floor by $h(x,t)$. The bathymetry $B(x)$ is the
negative depth of the ocean, i.e., $h(x,t)+B(x)=0$ when the
ocean is at rest.
Introducing the variable $v\coloneqq hu$, the shallow water equations can be written as
\begin{subequations}
	\label{eq:AP-PDE}
	\begin{alignat}{2}
	h_t+v_x&=0 \quad && \text{ on } \mathcal{D}\times [0,T_F],\label{eq:AP-PDE1}\\
	v_t+\left(\frac{v^2}{h}+\frac{1}{2}gh^2\right)_x+ghB_x&=0 && \text{ on
	} \mathcal{D}\times [0,T_F],\label{eq:AP-PDE2}\\
	h(x,0)=-B_0(x),\  v(x,0)&=0 && \text{ for } x\in \mathcal{D}, \label{eq:AP-PDE3}\\
	v(a,t)=v(b,t)&=0  && \text{ for } t\in [0,T_F],\label{eq:AP-PDE4}
	\end{alignat}
\end{subequations}
where $g$ is the gravitational constant and the subscripts $t,x$ denote
derivatives with respect to time and location. $B_0(x)$ is the reference bathymetry before any changes and is chosen based on the real topography \cite{tong2021extreme,fujiwara2011}.

The bathymetry $B(x)$, whose derivative enters in the right hand side
of \cref{eq:AP-PDE}, changes during an earthquake as a result of slip
between plates under the ocean floor. Since details of this slip
process are difficult to predict, we model the slip as a random
process, and thus also the bathymetry field $B$ is random. Since $B$
enters in the shallow water equations \cref{eq:AP-PDE}, the (space
and time-dependent) solutions $h$ and $v$ are random and hence also
the event objective we specify below is a random variable. Random
bathymetry changes are caused by slips under the earth between two
plates, which we propagate to the ocean floor using the Okada model
\cite{Okada85}.  Thus, we can denote the random parameter $\btheta$ in
this problem as these slips, with relation
\begin{equation}\label{eq:Okada->B}
B(x) =  B_0(x) + \sigma (O \btheta)(x) \: \text{ with } \:\btheta=(\theta_1,\ldots,\theta_{20})^\top \:\text{ and }\: (O\btheta)(x) \coloneqq \sum_{i=1}^{20}\theta_i O_i(x),
\end{equation}
where $O_i$ is the bathymetry change due to the $i$th slip patch, $\sigma^2=10$ and
$\btheta\sim \mathcal \NN(\bzero,\bI_{20})$.

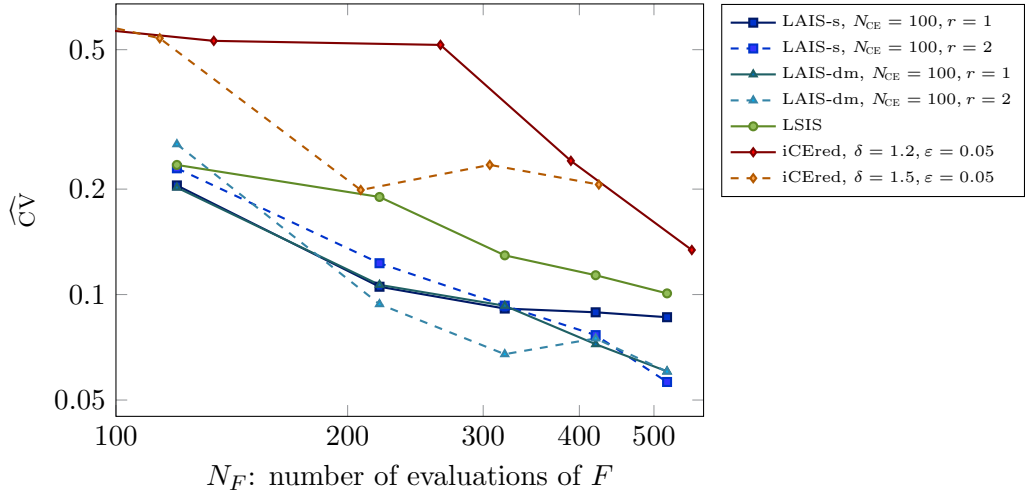
\begin{figure}[tb]
	\centering
	\begin{tikzpicture}[every mark/.append style={mark size=1.5pt}]
	\begin{axis}[width =  0.5\textwidth, 
	height =0.35\textwidth, 
	xmode=log, 
	ymode = log,
	scale only axis,  
	compat = 1.5,
	xlabel={$\NF$: number of evaluations of $F$},
	ylabel={$\cv$},
	legend style = {font=\tiny,nodes=right},
	legend pos=outer north east,
        xmin = 100,xmax=580,
	cycle list name= tsunami,
        xtick={100,200,300,400,500},
        xticklabels={100,200,300,400,500},
        ytick={0.05,0.1,0.2,0.5},
        yticklabels={0.05,0.1,0.2,0.5},
	]
		\addplot table[x expr=\thisrowno{0}+20,y=cv] {\data/tsunami_RMIS_single_nCE100_r1.txt};
	\addlegendentry{\RMIS-s, $\nCE=100, r=1$}
			\addplot table[x expr=\thisrowno{0}+20,y=cv] {\data/tsunami_RMIS_single_nCE100_r2.txt};
	\addlegendentry{\RMIS-s, $\nCE=100, r=2$}
			\addplot table[x expr=\thisrowno{0}+20,y=cv] {\data/tsunami_RMIS_DM_nCE100_r1.txt};
	\addlegendentry{\RMIS-dm, $\nCE=100, r=1$}
				\addplot table[x expr=\thisrowno{0}+20,y=cv] {\data/tsunami_RMIS_DM_nCE100_r2.txt};
	\addlegendentry{\RMIS-dm, $\nCE=100, r=2$}
		\addplot table[x expr=\thisrowno{0}+20,y=cv] {\data/tsunami_shift2.txt};
	\addlegendentry{\LSIS}
		\addplot table[x =N,y=cv] {\data/tsunami_CE_delta1.2_eps0.05.txt};
	\addlegendentry{\iCEred, $\delta=1.2, \varepsilon=0.05$}
			\addplot table[x =N,y=cv] {\data/tsunami_CE_delta1.5_eps0.05.txt};
	\addlegendentry{\iCEred, $\delta=1.5, \varepsilon=0.05$}
	\end{axis}
	\end{tikzpicture}
	\caption{Comparison of coefficient of variation for tsunami
          problem for different algorithms. As before, the comparison
          is with respect to the necessary number of evaluations of
          $F$, each of which requires the solution of a time-dependent
          1D shallow water equation.  }\label{fig:tsunami-N}
\end{figure}

The final observation is a smooth approximation of the maximal average
tsunami wave height in the interval $[c,d]$ on shore over the time period
$[0, T_F]$, i.e., 
\begin{equation}
\label{eq:AP-F regularized}
\begin{aligned}
  F(\btheta)
  \coloneqq
\gamma\ln\left[ 
\dfrac{1}{T_F}\int_{0}^{T_F}\exp \left(\dfrac{1}{\gamma}\fint_c^d(h+B_0)dx\right) dt\right],
\end{aligned}
\end{equation}
where we choose $\gamma=0.003$.  The event threshold is $z=0.5737$,
i.e., we are interested in estimating the probability that
$F(\btheta)\ge 0.5737$.  We compare the coefficient of variation from
an average of 10 independent runs of \RMIS, \LSIS and \iCEred for
different total number of evaluations of $F$ in \Cref{fig:tsunami-N}
and different subspace ranks $r = 1$ and $r=2$, which correspond to
$\epsilon=0.6$ and $\epsilon= 0.05$, respectively.
We observe that
\RMIS (both variants with standard and deterministic mixture weights)
achieve a lower CV compared to the other two methods, and also
performs robustly with respect to the numerical parameters in the
algorithm. While more comparisons will be necessary to compare these
algorithms rigorously for complex, PDE-governed problems, this
indicates the efficiency of \RMIS for estimating low probabilities in
complex systems.

\section{Conclusion and discussions}\label{sec:discussion}
We developed an adaptive importance sampling scheme for rare event
probability estimation in high dimensions. The approach (called \RMIS)
combines ideas from large deviation theory (LDT), importance sampling
(IS) based on cross-entropy (CE), and multiple importance sampling
(MIS). It consists of: (1) a sampling-free preprocessing step, namely
finding the LDT optimizer and using second-order derivative
information to identify an intrinsic low-dimensional subspace that is
most informative of the rare event probability. Additionally, the LDT
optimizer provides a good initial proposal for AIS; (2) an adaptive
sampling procedure: using CE to update a Gaussian proposal and MIS to
build an estimator that reuses all previous sample evaluations. Our
method improves the state-of-the-art \iCEred method proposed in \cite{uribe2021cross} in several
directions. \RMIS achieves a consistently smaller coefficient of variance and relative root-mean-square error compared
with \iCEred for the same number of evaluations of $F$,
since all samples in step (2) are used in the final estimator, and
since even
the initial set of samples are informative as they are
guided by the LDT optimizer. \RMIS requires no gradient evaluations in
step (2) and $O(10)$ ($<70$ in our examples) 
gradient
evaluations in step (1), while \iCEred requires gradient computation
for each sample.
In addition, \RMIS does not require tuning numerical parameters that
have significant influence on the performance of the algorithm.  A choice
one has to make is $\nCE$, i.e., how often one updates the biasing density,
but as long as $\nCE$ is moderately large ($\ge 200$), the performance
is robust with respect to this choice. 
Another numerical parameter in \RMIS is the rank of the
low-dimensional subspace, which is chosen based on the magnitude of
eigenvalues of $\HLDT$ (see \cref{sec:LDT-low}). In our experiments,
the performance of \RMIS is stable as long as we include the normal direction $\nabla F(\bthetastar)$.

There are opportunities to refine and extend \RMIS. First, a variant
of \RMIS with deterministic mixture weights that can be proven to be
bias-free would be desirable. A potential introduction of bias in
adaptive multiple importance samplers is, however, a known challenge \cite{cornuet2012adaptive}.
Further, the
parametric biasing distributions used in this paper are Gaussians,
which may suffer from degeneracy in high-dimensions for
problems without intrinsic low-rank structure. Thus, one could explore
the use of more general
parametric models such as the von Mises-Fisher-Nakagami mixture
\cite{papaioannou2019improved}, or a normalizing flow model from
machine learning \cite{kobyzev2020normalizing}.  Another extension is
to apply \RMIS for infinite-dimensional or non-Gaussian prior
distributions. For infinite-dimensional problem with intrinsic
low-rank structure, the main difficulty is to properly define the
weights and biasing densities.  
There are several challenges to extend \RMIS to non-Gaussian
distributions.
Firstly, the LDT rate function is 
no longer quadratic, and needs to be derived or even implicitly used in the form of the Legendre transform of the cumulant
generating function (some examples can be found in
\cite{tong2021extreme,tong2022optimization}) when solving the LDT optimization problem
\cref{eq:LDT-opt}.
In addition, both the local LIS and \Cref{thm:SO} are based on a Gaussian
prior, thus the technique of identifying a low-dimensional subspace in
\RMIS is no longer valid for non-Gaussian distributions. Certified
dimension reduction using $\HFIS$ \cite{zahm2018certified} may be a
possible alternative.

\section*{Acknowledgments}
We appreciate many helpful discussions with Eric Vanden-Eijnden and
Timo Schorlepp.

\bibliographystyle{siamplain}
\bibliography{sampling}

\begin{thebibliography}{10}

\bibitem{au2001estimation}
{\sc S.-K. Au and J.~L. Beck}, {\em Estimation of small failure probabilities
  in high dimensions by subset simulation}, Probabilistic Engineering
  Mechanics, 16 (2001), pp.~263--277.

\bibitem{bilmes1998gentle}
{\sc J.~A. Bilmes}, {\em A gentle tutorial of the {EM} algorithm and its
  application to parameter estimation for {G}aussian mixture and hidden
  {M}arkov models}, International Computer Science Institute, 4 (1998), p.~126.

\bibitem{bucklew2007introduction}
{\sc J.~Bucklew}, {\em Introduction to rare event simulation}, Springer, 2007.

\bibitem{bugallo2017adaptive}
{\sc M.~F. Bugallo, V.~Elvira, L.~Martino, D.~Luengo, J.~Miguez, and P.~M.
  Djuric}, {\em Adaptive importance sampling: The past, the present, and the
  future}, IEEE Signal Processing Magazine, 34 (2017), pp.~60--79.

\bibitem{cappe2004population}
{\sc O.~Capp{\'e}, A.~Guillin, J.-M. Marin, and C.~P. Robert}, {\em Population
  {M}onte {C}arlo}, Journal of Computational and Graphical Statistics, 13
  (2004), pp.~907--929.

\bibitem{cerou2012sequential}
{\sc F.~C{\'e}rou, P.~Del~Moral, T.~Furon, and A.~Guyader}, {\em Sequential
  {M}onte {C}arlo for rare event estimation}, Statistics and computing, 22
  (2012), pp.~795--808.

\bibitem{cornuet2012adaptive}
{\sc J.-M. Cornuet, J.-M. Marin, A.~Mira, and C.~P. Robert}, {\em Adaptive
  multiple importance sampling}, Scandinavian Journal of Statistics, 39 (2012),
  pp.~798--812.

\bibitem{cui2014likelihood}
{\sc T.~Cui, J.~Martin, Y.~M. Marzouk, A.~Solonen, and A.~Spantini}, {\em
  Likelihood-informed dimension reduction for nonlinear inverse problems},
  Inverse Problems, 30 (2014), p.~114015.

\bibitem{CullumWilloughby85}
{\sc J.~K. Cullum and R.~A. Willoughby}, {\em Lanczos Algorithms for Large
  Symmetric Eigenvalue Computations, Vol. 1: Theory}, Progress in Scientific
  Computing, Birkh{\"a}user-Verlag, Boston, Basel, Berlin, 1985.

\bibitem{de2005tutorial}
{\sc P.-T. De~Boer, D.~P. Kroese, S.~Mannor, and R.~Y. Rubinstein}, {\em A
  tutorial on the cross-entropy method}, Annals of Operations Research, 134
  (2005), pp.~19--67.

\bibitem{dematteis2019extreme}
{\sc G.~Dematteis, T.~Grafke, and E.~Vanden-Eijnden}, {\em Extreme event
  quantification in dynamical systems with random components}, SIAM/ASA Journal
  on Uncertainty Quantification, 7 (2019), pp.~1029--1059.

\bibitem{dembo1998large}
{\sc A.~Dembo and O.~Zeitouni}, {\em Large Deviations Techniques and
  Applications}, Applications of Mathematics, Springer, 1998.

\bibitem{ditlevsen1996structural}
{\sc O.~Ditlevsen and H.~O. Madsen}, {\em Structural reliability methods},
  vol.~178, Wiley New York, 1996.

\bibitem{ditlevsen1990general}
{\sc O.~Ditlevsen, R.~E. Melchers, and H.~Gluver}, {\em General
  multi-dimensional probability integration by directional simulation},
  Computers \& Structures, 36 (1990), pp.~355--368.

\bibitem{du2001most}
{\sc X.~Du and W.~Chen}, {\em A most probable point-based method for efficient
  uncertainty analysis}, Journal of Design and Manufacturing Automation, 4
  (2001), pp.~47--66.

\bibitem{ebener2019instanton}
{\sc L.~Ebener, G.~Margazoglou, J.~Friedrich, L.~Biferale, and R.~Grauer}, {\em
  Instanton based importance sampling for rare events in stochastic {PDE}s},
  Chaos: An Interdisciplinary Journal of Nonlinear Science, 29 (2019),
  p.~063102.

\bibitem{fujiwara2011}
{\sc T.~Fujiwara, S.~Kodaira, T.~No, Y.~Kaiho, N.~Takahashi, and Y.~Kaneda},
  {\em The 2011 {T}ohoku-{O}ki earthquake: Displacement reaching the trench
  axis}, Science, 334 (2011), pp.~1240--1240.

\bibitem{geyer2019cross}
{\sc S.~Geyer, I.~Papaioannou, and D.~Straub}, {\em Cross entropy-based
  importance sampling using {G}aussian densities revisited}, Structural Safety,
  76 (2019), pp.~15--27.

\bibitem{grafke2019numerical}
{\sc T.~Grafke and E.~Vanden-Eijnden}, {\em Numerical computation of rare
  events via large deviation theory}, Chaos: An Interdisciplinary Journal of
  Nonlinear Science, 29 (2019), p.~063118.

\bibitem{HalkoMartinssonTropp11}
{\sc N.~Halko, P.~G. Martinsson, and J.~A. Tropp}, {\em Finding structure with
  randomness: {P}robabilistic algorithms for constructing approximate matrix
  decompositions}, SIAM Review, 53 (2011), pp.~217--288.

\bibitem{hesterberg1995weighted}
{\sc T.~Hesterberg}, {\em Weighted average importance sampling and defensive
  mixture distributions}, Technometrics, 37 (1995), pp.~185--194.

\bibitem{kahn1953methods}
{\sc H.~Kahn and A.~W. Marshall}, {\em Methods of reducing sample size in
  {M}onte {C}arlo computations}, Journal of the Operations Research Society of
  America, 1 (1953), pp.~263--278.

\bibitem{katafygiotis2008geometric}
{\sc L.~S. Katafygiotis and K.~M. Zuev}, {\em Geometric insight into the
  challenges of solving high-dimensional reliability problems}, Probabilistic
  Engineering Mechanics, 23 (2008), pp.~208--218.

\bibitem{kobyzev2020normalizing}
{\sc I.~Kobyzev, S.~J. Prince, and M.~A. Brubaker}, {\em Normalizing flows: An
  introduction and review of current methods}, IEEE transactions on pattern
  analysis and machine intelligence, 43 (2020), pp.~3964--3979.

\bibitem{lemaire2013structural}
{\sc M.~Lemaire}, {\em Structural reliability}, John Wiley \& Sons, 2013.

\bibitem{leveque2008high}
{\sc R.~J. LeVeque and D.~L. George}, {\em High-resolution finite volume
  methods for the shallow water equations with bathymetry and dry states}, in
  Advanced numerical models for simulating tsunami waves and runup, World
  Scientific, 2008, pp.~43--73.

\bibitem{liu2008monte}
{\sc J.~S. Liu}, {\em Monte {C}arlo strategies in scientific computing},
  Springer Science \& Business Media, 2008.

\bibitem{metropolis1949monte}
{\sc N.~Metropolis and S.~Ulam}, {\em The {M}onte {C}arlo method}, Journal of
  the American statistical association, 44 (1949), pp.~335--341.

\bibitem{Okada85}
{\sc Y.~Okada}, {\em Surface deformation due to shear and tensile faults in a
  half-space}, Bulletin of the seismological society of America, 75 (1985),
  pp.~1135--1154.

\bibitem{owen2000safe}
{\sc A.~Owen and Y.~Zhou}, {\em Safe and effective importance sampling},
  Journal of the American Statistical Association, 95 (2000), pp.~135--143.

\bibitem{paananen2021implicitly}
{\sc T.~Paananen, J.~Piironen, P.-C. B{\"u}rkner, and A.~Vehtari}, {\em
  Implicitly adaptive importance sampling}, Statistics and Computing, 31
  (2021), pp.~1--19.

\bibitem{papaioannou2019improved}
{\sc I.~Papaioannou, S.~Geyer, and D.~Straub}, {\em Improved cross
  entropy-based importance sampling with a flexible mixture model}, Reliability
  Engineering \& System Safety, 191 (2019), p.~106564.

\bibitem{press2007numerical}
{\sc W.~H. Press, S.~A. Teukolsky, W.~T. Vetterling, and B.~P. Flannery}, {\em
  Numerical recipes 3rd edition: The art of scientific computing}, Cambridge
  university press, 2007.

\bibitem{rackwitz2001reliability}
{\sc R.~Rackwitz}, {\em Reliability analysis -- a review and some
  perspectives}, Structural Safety, 23 (2001), pp.~365--395.

\bibitem{rubinstein1997optimization}
{\sc R.~Y. Rubinstein}, {\em Optimization of computer simulation models with
  rare events}, European Journal of Operational Research, 99 (1997),
  pp.~89--112.

\bibitem{sapsis2018new}
{\sc T.~P. Sapsis}, {\em New perspectives for the prediction and statistical
  quantification of extreme events in high-dimensional dynamical systems},
  Philosophical Transactions of the Royal Society A: Mathematical, Physical and
  Engineering Sciences, 376 (2018), p.~20170133.

\bibitem{schorlepp2022spontaneous}
{\sc T.~Schorlepp, T.~Grafke, S.~May, and R.~Grauer}, {\em Spontaneous symmetry
  breaking for extreme vorticity and strain in the three-dimensional
  {N}avier--{S}tokes equations}, Philosophical Transactions of the Royal
  Society A, 380 (2022), p.~20210051.

\bibitem{schueller1987critical}
{\sc G.~I. Schu{\"e}ller and R.~Stix}, {\em A critical appraisal of methods to
  determine failure probabilities}, Structural Safety, 4 (1987), pp.~293--309.

\bibitem{tong2022optimization}
{\sc S.~Tong, A.~Subramanyam, and V.~Rao}, {\em Optimization under rare chance
  constraints}, SIAM Journal on Optimization, 32 (2022), pp.~930--958.

\bibitem{tong2021extreme}
{\sc S.~Tong, E.~Vanden-Eijnden, and G.~Stadler}, {\em Extreme event
  probability estimation using {PDE}-constrained optimization and large
  deviation theory, with application to tsunamis}, Communications in Applied
  Mathematics and Computational Science, 16 (2021), pp.~181--225.

\bibitem{ullmann2015multilevel}
{\sc E.~Ullmann and I.~Papaioannou}, {\em Multilevel estimation of rare
  events}, SIAM/ASA Journal on Uncertainty Quantification, 3 (2015),
  pp.~922--953.

\bibitem{uribe2021cross}
{\sc F.~Uribe, I.~Papaioannou, Y.~M. Marzouk, and D.~Straub}, {\em
  Cross-entropy-based importance sampling with failure-informed dimension
  reduction for rare event simulation}, SIAM/ASA Journal on Uncertainty
  Quantification, 9 (2021), pp.~818--847.

\bibitem{varadhan1984large}
{\sc S.~R.~S. Varadhan}, {\em Large deviations and applications}, vol.~46,
  SIAM, 1984.

\bibitem{veach1995optimally}
{\sc E.~Veach and L.~J. Guibas}, {\em Optimally combining sampling techniques
  for {M}onte {C}arlo rendering}, in Proceedings of the 22nd annual conference
  on Computer graphics and interactive techniques, 1995, pp.~419--428.

\bibitem{wagner2020multilevel}
{\sc F.~Wagner, J.~Latz, I.~Papaioannou, and E.~Ullmann}, {\em Multilevel
  sequential importance sampling for rare event estimation}, SIAM Journal on
  Scientific Computing, 42 (2020), pp.~A2062--A2087.

\bibitem{wahalbimc}
{\sc S.~Wahal and G.~Biros}, {\em {BIMC}: {T}he {B}ayesian inverse {M}onte
  {C}arlo method for goal-oriented uncertainty quantification. {P}art {I}},
  arXiv preprint arXiv:1911.00619,  (2019).

\bibitem{zahm2018certified}
{\sc O.~Zahm, T.~Cui, K.~Law, A.~Spantini, and Y.~Marzouk}, {\em Certified
  dimension reduction in nonlinear {B}ayesian inverse problems}, Mathematics of
  Computation, 91 (2022), pp.~1789--1835.

\end{thebibliography}
\end{document}